\documentclass[10pt, conference, letterpaper]{IEEEtran}
\IEEEoverridecommandlockouts
\makeatletter
\def\ps@headings{%
\def\@oddhead{\mbox{}\scriptsize\rightmark \hfil \thepage}%
\def\@evenhead{\scriptsize\thepage \hfil \leftmark\mbox{}}%
\def\@oddfoot{}%
\def\@evenfoot{}}
\makeatother
\usepackage{subfigure}
\usepackage{colortbl}
\usepackage{bm}
\usepackage{cite}
\usepackage{exscale}
\usepackage{relsize}

\usepackage{graphicx}  
\usepackage{url}       

\usepackage{amsthm,amsmath,amssymb}   
\usepackage{mathrsfs}
\usepackage{cite}

\usepackage{amsfonts,amssymb}

\usepackage{stfloats}

\usepackage{cases}
\usepackage{algorithm}
\usepackage{multirow}
\usepackage{algorithmic}
\usepackage{stfloats}
\usepackage{colortbl,booktabs}
\usepackage{color}
\usepackage{epstopdf}

\newcommand{\ls}[1]
    {\dimen0=\fontdimen6\the\font
     \lineskip=#1\dimen0
     \advance\lineskip.5\fontdimen5\the\font
     \advance\lineskip-\dimen0
     \lineskiplimit=.9\lineskip
     \baselineskip=\lineskip
     \advance\baselineskip\dimen0
     \normallineskip\lineskip
     \normallineskiplimit\lineskiplimit
     \normalbaselineskip\baselineskip
     \ignorespaces
    }

\usepackage[absolute,showboxes]{textpos}

\TPMargin{3pt}

\begin{document}

\newcommand{\copyrightstatement}{
    \begin{textblock}{0.84}(0.08,0.93) 
         \noindent
         \footnotesize
         \copyright 2021 IEEE. Personal use of this material is permitted. Permission from IEEE must be obtained for all other uses, in any current or future media, including reprinting/republishing this material for advertising or promotional purposes, creating new collective works, for resale or redistribution to servers or lists, or reuse of any copyrighted component of this work in other works. DOI: 10.1109/JSAC.2021.3088684.
    \end{textblock}
}
\copyrightstatement

\title{Resource Allocation for 5G-UAV Based\\ Emergency Wireless Communications}
\author{\IEEEauthorblockN{Zhuohui Yao,~\IEEEmembership{Student Member,~IEEE}, Wenchi Cheng,~\IEEEmembership{Senior Member,~IEEE}, \\ Wei Zhang,~\IEEEmembership{Fellow,~IEEE}, and Hailin Zhang,~\IEEEmembership{Member,~IEEE}}~\\[0.2cm]
\vspace{-20pt}

\thanks{\ls{0.5}This work was supported in part by the National Natural Science Foundation of China (Nos. 61771368 and 61671347), the Young Elite Scientists Sponsorship Program by CAST (2016QNRC001), the Australian Research Councils Project funding scheme under LP160100672, and Shenzhen Science
\& Innovation Fund under Grant JCYJ20180507182451820.

Z. Yao, W. Cheng, and H. Zhang are with the State Key Laboratory of Integrated Services Networks, Xidian University, Xian 710071, China (e-mail: zhhyao\@stu.xidian.edu.cn; wccheng\@xidian.edu.cn; hlzhang@xidian.edu.cn).

W. Zhang is with the School of Electrical Engineering and Telecommunications, The University of New South Wales, Sydney, Australia
(e-mail: w.zhang\@unsw.edu.au).
}
}

\maketitle

\begin{abstract}
For unforeseen natural disasters, such as earthquakes, hurricanes, and floods, etc., the traditional communication infrastructure is unavailable or seriously disrupted along with persistent secondary disasters. Under such circumstances, it is highly demanded to deploy emergency wireless communication (EWC) networks to restore connectivity in accident/incident areas. The emerging fifth-generation (5G)/beyond-5G (B5G) wireless communication system, like unmanned aerial vehicle (UAV) assisted networks and intelligent reflecting surface (IRS) based communication systems, are expected to be designed or re-farmed for supporting temporary high quality communications in post-disaster areas. However, the channel characteristics of post-disaster areas quickly change as the secondary disaster resulted topographical changes, imposing new but critical challenges for EWC networks. In this paper, we propose a novel heterogeneous $\mathcal{F}$ composite fading channel model for EWC networks which accurately models and characterizes the composite fading channel with reflectors, path-loss exponent, fading, and shadowing parameters in 5G-UAV based EWC networks. Based on the model, we develop the optimal power allocation scheme with the simple closed-form expression and the numerical results based optimal joint bandwidth-power allocation scheme. We derive the corresponding capacities and compare the energy efficiency between IRS and traditional relay based 5G-UAVs. Numerical results show that the new heterogeneous Fisher-Snedecor $\mathcal{F}$ composite fading channel adapted resource allocation schemes can achieve higher capacity and energy efficiency than those of traditional channel model adapted resource allocation schemes, thus providing better communications service for post-disaster areas.
\end{abstract}

\vspace{10pt}


\section{Introduction}

In recent years, a number of disasters, such as Mexico earthquake in 2017, super typhoon Kong-rey, and hurricane Laura in 2020, etc., severely impact the normal life of human beings. For quickly and efficiently recovering after disaster, emergency communication, which provides a specified but agile communication mechanism for specified coverage, high capacity, low latency, and high energy efficiency in emergency situations, is very important in post-disaster area and persistently requires novel wireless communications techniques for rescue capabilities enhancement~\cite{7439936}~\cite{2016A}. On the other hand, with rapid development of wireless communications, such as the gradually matured fourth generation (4G) wireless networks and eventually implemented fifth generation (5G)/beyond-5G wireless networks, a number of techniques have been developed for extensive coverage, high capacity, and high energy efficiency~\cite{6235944}~\cite{7306369}. Among them, the unmanned aerial vehicle (UAV)-based transmission, is an attractive scheme that can be utilized to quickly establish the temporary communication networks, which is very useful for communications in post-disaster areas.

In typical emergency wireless communication (EWC) networks, due to their flexibility and 3D movement capability, UAVs are often placed as relays or temporary base stations to satisfy the communication needs in emergency saving scenarios~\cite{UAV_relay2}~\cite{UAV-aided}. In~\cite{ZhangR_2016TCOM}, the authors optimized the trajectory of UAVs, which are relay base stations, to achieve the maximum throughput. However, there are some typical shortcomings of UAVs as relays or temporary base stations such as unclear channel model and limited resources for UAV based EWC networks, which results in low capacity and energy efficiency for UAVs based communications scenarios.


For channel model, since the channel characteristics of EWC networks vary along with the unpredictable secondary disasters resulted topography complexly changes, the general fading distributions, of which include Rayleigh, Rician, and Nakagami-$m$ channel, are not applicable to the EWC networks. For EWC networks, the variation of shadowing caused by secondary disasters severely impacts the channel model~\cite{TCOM_2020, TCOM_18, TCOM_19, TCOM_20}. Several composite fading-shadowing models, which are mixtures of the inverse gamma or inverse Nakagami-$m$ shadowing with a classical multi-path fading distribution, have been proposed~\cite{682805, 8625524, 8474223} to model the impact of both fading and shadowing on wireless links. Among the composite models, the Fisher-Snedecor $\mathcal{F}$ distribution, which considers Nakagami-$m$ fading mixed with inverse Nakagami-$m$ shadowing, has received much attentions~\cite{7886273, 8458088,8371018}. It outperforms other general composite fading models in most cases with providing a better fit to the actual data. It can also efficiently show the characteristics of wireless channel with composite fading and shadowing. For EWC networks, the link between the on-site command center and 5G-UAV undergoes the large scale fading while the link between 5G-UAV and trapped users experiences the Fisher-Snedecor $\mathcal{F}$ distribution. Therefore, an accurate channel model is highly demanded for such a complex communication scenario.



For capacity and energy efficiency, the emerging high-volume traffic in post-disaster areas results in the communication congestion and amounts of resources demand, which imposes challenges for performance enhancement. Based on the shadowed-Rician fading channel, the authors of~\cite{UAV_CL} proposed an UAV-enabled relaying system with energy harvesting functionality while the authors of~\cite{UAV_energy_relay} studied the energy-efficient UAV communications to support ground nodes. In~\cite{TVT_self_ennergized}, the authors presented a unified energy management framework by resorting to wireless power transfer, simultaneous wireless information and power transfer, and self-interference energy harvesting schemes in multi-UAV relay networks. To further increase capacity and energy efficiency, intelligent reflective surface (IRS), which is easy to be deployed and smartly reconfigure the wireless propagation environments, is expected to replace the traditional relays with its very low energy consumption~\cite{IRS_zhang_access}. In recent years, IRS-enabled applications for increasing capacity and energy efficiency under different communication systems have been investigated~\cite{9174910, 8981888, 9110888, 9179528, 9149411, 9162097}. For EWC scenarios, it is promising and perspective to exploit the advanced IRS-assisted 5G-UAV communication systems to increase capacity and energy efficiency under the complex channel conditions. However, considering the multiple factors such as the path-loss, fading, and shadowing parameters, until now it is still unclear how to allocate the limited resources to enhance capacity and energy efficiency under the complex channel conditions.

Motivated by the accurate channel modeling and performance enhancement requirements for EWC networks, we consider an emergency communication architecture where multiple 5G-UAVs are employed with IRSs to reflect and aggregate the transmit signals from the on-site command center (the location of on-site command center is selected with large scale fading between the on-site command center and 5G-UAVs to facilitate post-disaster rescue~\cite{7439936}~\cite{2016A}) to the destination in the trapped areas though complex EWC channels. First, a novel heterogeneous Fisher-Snedecor $\mathcal{F}$ composite fading channel model with reflectors, path-loss exponent, fading, and shadowing parameters for the emergency wireless communication systems is given. Then, the power allocation and joint bandwidth-power optimization schemes, which are adaptive to the heterogeneous $\mathcal{F}$ composite fading model, are proposed to maximize the capacity and energy efficiency of 5G-UAVs based EWC networks. We compare the capacities under the cases of heterogeneous $\mathcal{F}$ composite fading channel based and general fading channel based resource allocation schemes. We also show the enhancement of energy efficiency with IRS integrated 5G-UAVs. Numerical results show that the new heterogeneous Fisher-Snedecor $\mathcal{F}$ composite fading channel adapted resource allocation schemes can achieve higher capacity and energy efficiency than those of traditional channel model adapted resource allocation schemes, thus providing better communications service for post-disaster areas.

The rest of this paper is organized as follows. Section~\ref{sec:sys} describes the multiple 5G-UAV based EWC network system model. Section~\ref{optimization} presents the resource allocation schemes over the heterogeneous $\mathcal{F}$ composite fading channel to enhance capacity and energy efficiency performances for EWC networks. Section~\ref{sec:simu} numerically evaluates the capacities and energy efficiencies under our proposed optimal power allocation scheme and joint bandwidth-power allocation scheme for the heterogeneous $\mathcal{F}$ composite fading channel, respectively. Finally, Section~\ref{sec:conc} concludes this paper.


\section{The 5G-UAV based EWC System Model}\label{sec:sys}

\begin{figure*}
\centering
\vspace{-2pt}
\includegraphics[scale=0.55]{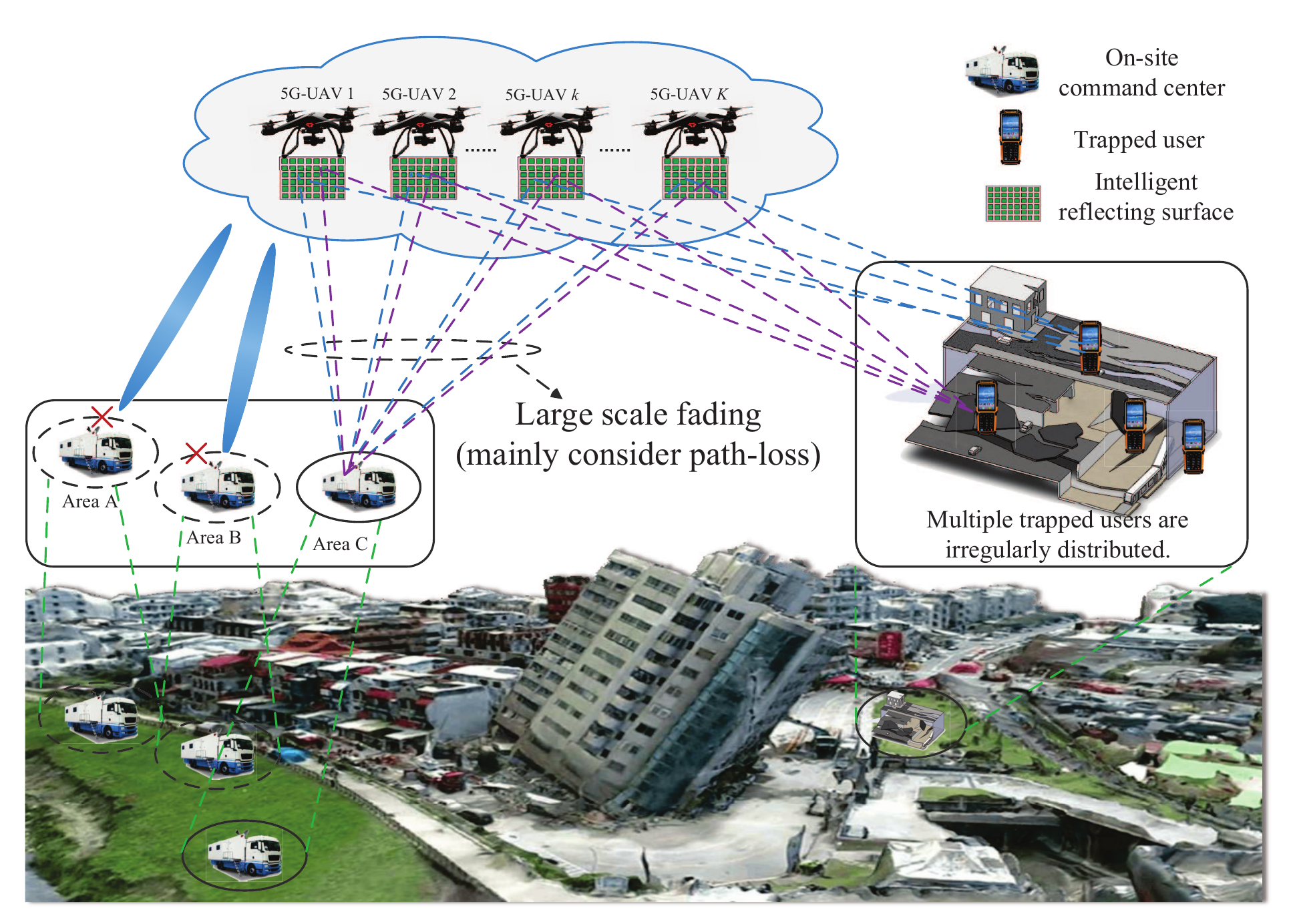}
\caption{5G-UAV based emergency wireless communications networks.} \label{5G-UAV-system_model}
\vspace{-8pt}
\end{figure*}

Figure~\ref{5G-UAV-system_model} shows an emergency wireless communication system that includes the on-site command center, $K$ 5G-UAVs integrated with IRSs (each consisting of $N$ passive reflector elements), and the irregularly distributed trapped users in the post-disaster area. Under such an emergency wireless communication scenario, there is only one on-site command center for sending/receiving rescuing instructions within one local post-disaster area to serve nearby trapped users. Consequently, the location of on-site command center, which plays an important role in the rescue process, is reasonably selected to facilitate communications with the trapped users in the post-disaster area (for example, Area C is selected for the on-site command center in Fig.~\ref{5G-UAV-system_model}). For the emergency wireless communication system, we consider an on-site command center transmits command information to multiple trapped users. Moreover, it is very important to accurately and timely communicate with the trapped users in the post-disaster area. Frequency-division and time-division, as common but effective ways, are adopted for trapped users in post-disaster areas. Therefore, each trapped user is connected to the on-site command center with its own-assigned frequency-band/time-slot and can be individually considered as an optimization problem. Since the topography complexly changes, the signal, which is transmitted from the on-site command center to the trapped user, will suffer severe fading and shadowing, resulting in the communication outage. Therefore, we employ the 5G-UAVs integrated with IRSs, which are hovering between the on-site command center and the trapped user, to reflect the transmit signal with the optimal phase shift\footnote{We assume that IRSs on 5G-UAVs can be adaptive to the channel, thus achieving the optimal beamforming~\cite{ZhangR_magazine}.}.

For the whole system considered, the on-site command center sends a signal to the trapped user. The available bandwidth is divided into $K$ sub-channels for 5G-UAVs. The signal received by the trapped user is the superposition of $K$-path signals, denoted by $y$, represented as follows:
\begin{eqnarray}
y\!=\!\bigg[\sum\limits_{k=1}^{K}\sum\limits_{n=1}^{N}\sqrt{(L_{k}^{SR})^{-\alpha}(L_{k}^{RD})^{-\alpha}g_{k,n}}e^{-j\phi_{k,n}} \bigg] x \!+\! \omega,
\label{eq:multi-uav_rms}
\end{eqnarray}
\\
where $x$ represents the transmit signal from the on-site command center, $\phi_{k,n}$ is the phase shift of the $n$-th reflecting element of IRS, and $\omega \sim \mathcal{C}\mathcal{N}(0,\delta^{2})$ is the additive white Gaussian noise (AWGN). $\sqrt{(L_{k}^{SR})^{-\alpha}}$ is the channel gain of the $k$-th sub-channel from the on-site command center to the $k$-th 5G-UAV with distance $L_{k}^{SR}$ and the path-loss exponent $\alpha$.
This is because the line-of-sight communication between the on-site command center and 5G-UAVs undergo the large scale fading, which mainly includes path loss and shadowing. Moreover, under such an emergency wireless communication scenario, the on-site command center and 5G-UAVs are moved to locate with a good propagation environment for facilitating communications. The shadowing can be effectively reduced due to the deployment of on-site command center and 5G-UAVs. Therefore, in this system, the links between the on-site command center and 5G-UAVs mainly undergo path loss. Following that, since the trapped users are located in the affected areas for post-disasters, the characteristics of channel from the 5G-UAVs to the trapped user vary along with the unpredictable secondary disasters resulted topography complexly changes in the emergency wireless communication networks, especially fading and shadowing. In view of such complex channel conditions, the links between 5G-UAVs and the trapped user experience the Fisher-Snedecor $\mathcal{F}$ distribution~\cite{7886273}, which is proposed as an accurate and tractable composite model to capture the multi-path fading and shadowing conditions in practical wireless propagation scenarios. Also, it outperforms the generalized-K composite fading model in most cases with providing a better fit to the actual data.
Therefore, $\sqrt{(L_{k}^{RD})^{-\alpha}g_{k,n}}$ is the channel gain of the $k$-th sub-channel from the 5G-UAV to the trapped user with distance $L_{k}^{RD}$, the path-loss exponent $\alpha$, and the Fisher-Snedecor $\mathcal{F}$ random variables (RVs) $g_{k,n}$. The RV $g_{k,n}$ is modelled as independently distributed with the probability density function (PDF), denoted by $f_{k,n}(g_{k,n})$, as follows:
\begin{align}
f_{k,n}(g_{k\!,n})\!\!=\!\!\frac{m_{k,n}^{m_{k,n}}(m_{s_{k,n}}\bar{g}_{k,n})^{m_{s_{k,n}}}g_{k,n}^{m_{k,n}-1}}{B(m_{k\!,n}\!, m_{s_{k\!,n}})\!(m_{k\!,n}g_{k\!,n}\!\!+\!\!m_{s_{k\!,n}}\bar{g}_{k\!,n})^{m_{k\!,n}\!+\!m_{s_{k\!,n}}}},
\label{eq:fgkn-pdf}
\end{align}
where $m_{k,n}$ and $m_{s_{k,n}}$ represent the fading and shadowing parameters, respectively, of the $n$-th RV at the $k$-th sub-channel, $\bar{g}_{k,n}=\mathbb{E}(g_{k,n})$ is the mean power, $B(a_{1}, a_{2})$ is the beta function with $B(a_{1}, a_{2})=\frac{\Gamma(a_{1})\Gamma(a_{2})}{\Gamma(a_{1}+a_{2})}$~\cite[Eq.(8. 384.1)]{2007859}, where $\Gamma(\varphi)=\int_{0}^{\infty}t^{\varphi-1}e^{-\varphi}\mathrm{d}t$ is the gamma function~\cite[Eq.(8.310)]{2007859}.

Based on Eq.~\eqref{eq:multi-uav_rms}, the instantaneous signal-to-noise ratio (SNR) corresponding to the $k$-th sub-channel, denoted by $\gamma_{k}$, is derived as follows
\begin{eqnarray}
\begin{aligned}
\gamma_{k}\!\!=\!\!\sum\limits_{n=1}^{N}(L_{k}^{SR})^{-\alpha}(L_{k}^{RD})^{-\alpha}g_{k,n} \frac{P(g_{k})}{\delta^{2}}\!=\!\frac{P(g_{k})g_{k}}{\delta^{2}L_{k}^{\alpha}},
\label{eq:ins_gamma}
\end{aligned}
\end{eqnarray}
where $L_{k}^{-\alpha}=\left(L_{k}^{SR}L_{k}^{RD}\right)^{-\alpha}$ is the path-loss from the on-site command center to the trapped user over the $k$-th sub-channel, $g_{k}=\sum\limits_{n=1}^{N}g_{k,n}$ is the sum power gain of $N$ reflectors, $\delta^{2}=N_{0}B$ is the variance of AWGN with the corresponding bandwidth $B$, and $P(g_{k})$ is the transmit power corresponding to the $k$-th sub-channel.

In our proposed emergency wireless communication scenario, the $K$ 5G-UAVs cooperate to aggregate the transmit signal for maximizing the SNR. Thus, we define the heterogeneous $\mathcal{F}$ composite fading channel with power gain $h$, where $h=\sum_{k=1}^{K}L_{k}^{-\alpha}g_{k}$.

\section{Resource Allocation Schemes \\for Multiple 5G-UAV Based EWC Networks}\label{optimization}

In this section, we derive the heterogeneous $\mathcal{F}$ composite fading channel model for 5G-UAV enabled EWC networks. We also show the impacts of channel parameters and number of 5G-UAVs on the PDF for power gain. Then, we develop the optimal power allocation scheme, which is adaptive to the heterogeneous $\mathcal{F}$ composite fading channel, to obtain the maximum ergodic capacity for 5G-UAV based EWC networks. We analyze the changes of the optimal power allocation scheme with the channel parameters of power gain, fading, and shadowing. Furthermore, the joint bandwidth-power optimization algorithm for maximizing the ergodic capacity of heterogeneous $\mathcal{F}$ composite fading channel is developed with its convergence revealed. Finally, the energy efficiencies of IRS-based 5G-UAV and relay-based 5G-UAV are analyzed.

\subsection{The Heterogeneous $\mathcal{F}$ Composite Fading Channel}
Based on the system model, we define the heterogeneous $\mathcal{F}$ composite fading channel for 5G-UAV enabled EWC networks. Following that, we derive the PDF of power gain $h$, and show the impacts of channel parameters and number of reflectors with IRS-assisted 5G-UAVs. Here, we first derive the PDF of power gain at the $k$-th sub-channel $h_{k}$ and the heterogeneous $\mathcal{F}$ composite fading channel $h$, respectively, as shown in the following.

\emph{Lemma 1}: For $h_{k}=L_{k}^{-\alpha}g_{k}$, the PDF of the $k$-th sub-channel power gain $h_{k}$, denoted by $f_{k}(h_{k})$, is
\begin{eqnarray}
\begin{aligned}
f_{k}(h_{k})\!=\!\frac{L_{k}^{\alpha}\Lambda_{k}}{\Gamma(Nm_{k}) \Gamma\left(N m_{s_{k}}\right)} G_{2,2}^{1,2}\left(\Lambda_{k}h_{k} \bigg| \! \begin{array}{c}
-Nm_{s_{k}},0 \\
N m_{k}\!-\!1,0
\end{array}\!\!\right),
\end{aligned}
\end{eqnarray}
where $m_{k}$ and $m_{s_{k}}$ represent the fading and shadowing parameters of the $k$-th sub-channel, respectively. We define $\Lambda_{k}=\frac{m_{k}L_{k}^{\alpha}}{Nm_{s_{k}} \bar{h}_{k}}$ for simplicity.

\emph{Proof}: The proof is provided in Appendix A. $\hfill\blacksquare$

\emph{Lemma 2}: For $h=\sum_{k=1}^{K}h_{k}$, the PDF of the heterogeneous $\mathcal{F}$ composite fading channel power gain $h$, denoted by $f(h)$, is
\begin{eqnarray}
\begin{aligned}
f(h)\!\!=\!\!\frac{\frac{\Lambda }{K}\prod_{k=1}^{K}L_{k}^{\alpha}}{\Gamma(KNm)\Gamma(KNm_{s})} G_{2,2}^{1,2}\left(\!\!\dfrac{\Lambda h}{K} \bigg|\!\! \begin{array}{c}
-KNm_{s},0 \\
KNm-1,0
\end{array}\!\!\right),
\end{aligned}\label{eq:_fh}
\end{eqnarray}
where $m$ and $m_{s}$ represent the average fading and shadowing parameters of the heterogeneous $\mathcal{F}$ composite fading channel, respectively. We define $\Lambda=\frac{m \bar{L}^{\alpha}}{Nm_{s} \bar{h}}$ for simplicity.

\emph{Proof}: The proof is provided in Appendix B. $\hfill\blacksquare$

\begin{figure}[h]
\centering
\includegraphics[scale=0.5]{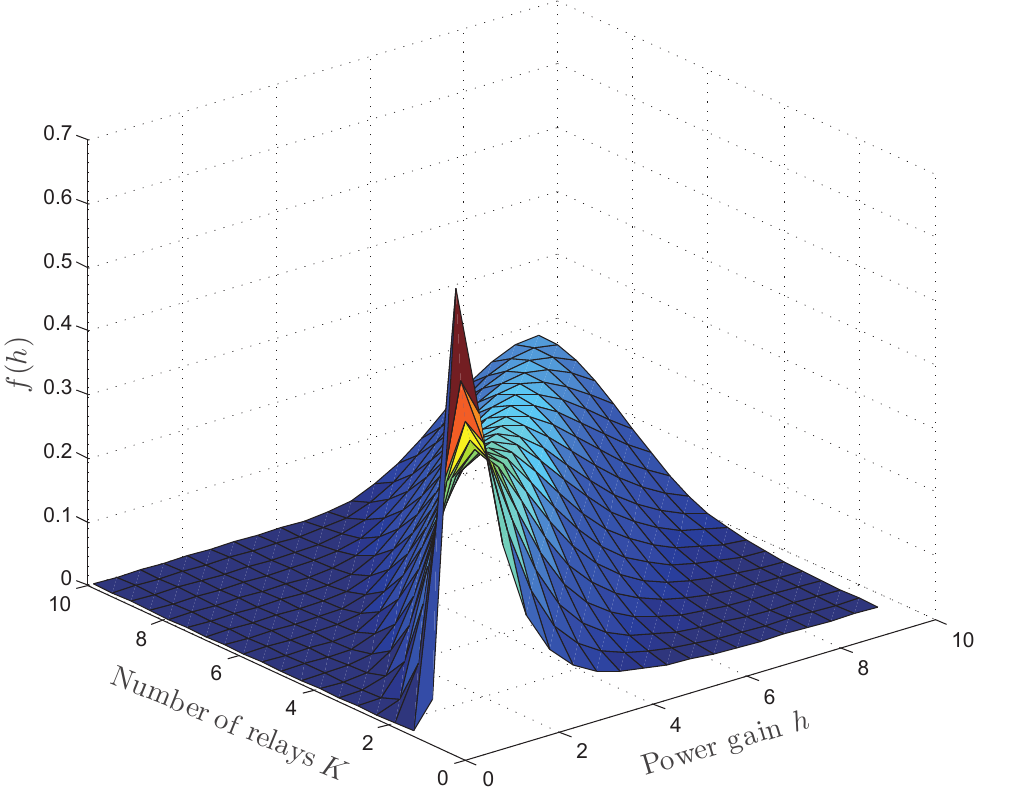}
\caption{The PDF of heterogeneous $\mathcal{F}$ composite fading channel versus the channel characteristics of power gain $h$ and the number of 5G-UAVs $K$.} \label{figure-pdf-K_h}
\end{figure}
\begin{figure}[h]
\centering
\includegraphics[scale=0.5]{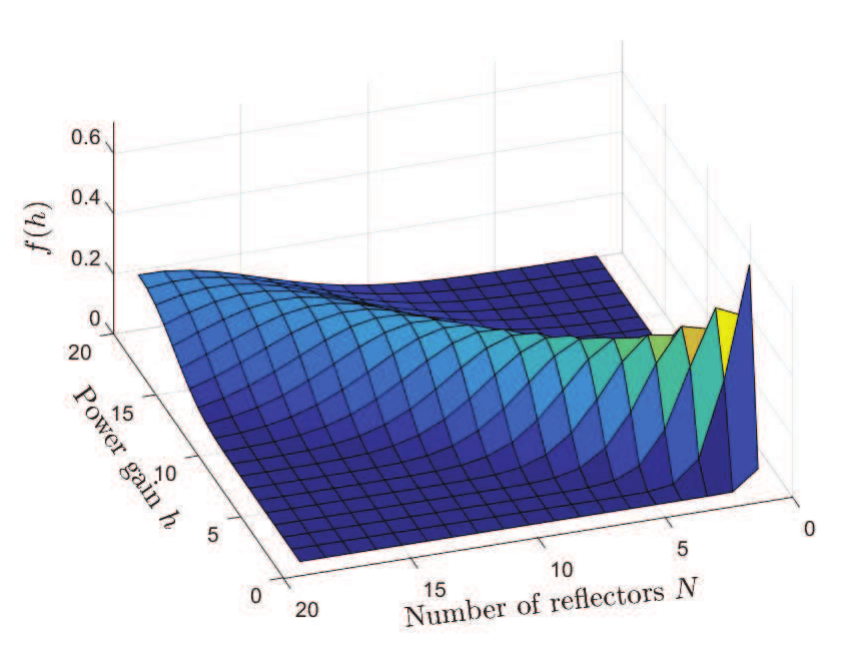}
\caption{The PDF of heterogeneous $\mathcal{F}$ composite fading channel versus the channel characteristics of power gain $h$ and the number of reflectors $N$.} \label{figure-pdf-N_h}
\end{figure}
\begin{figure}[h]
\centering
\includegraphics[scale=0.45]{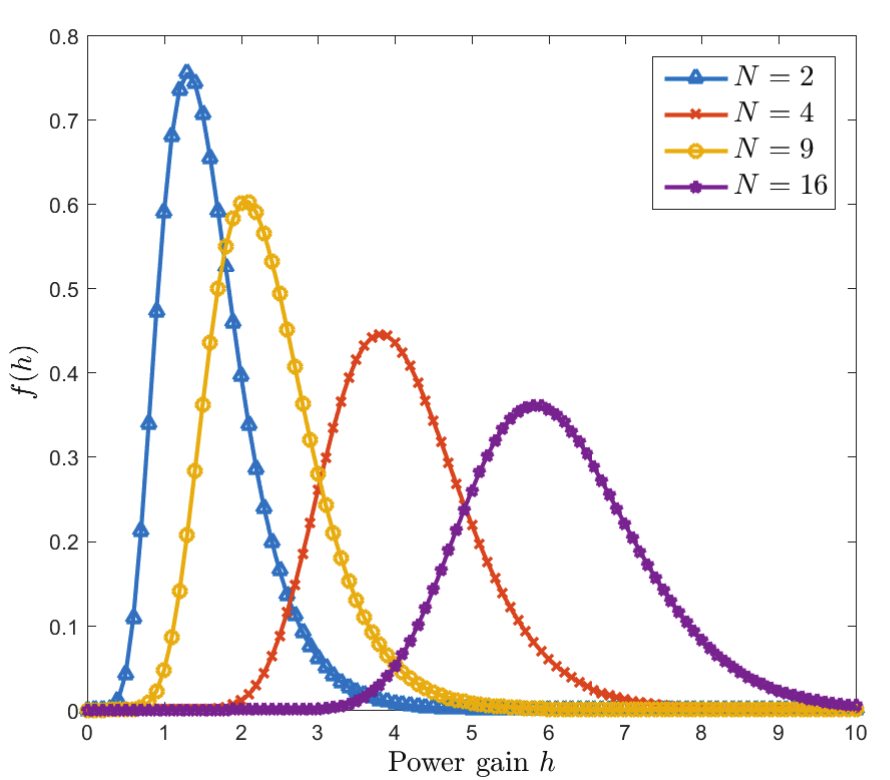}
\caption{The PDF of heterogeneous $\mathcal{F}$ composite fading channel versus the power gain $h$.} \label{pdf-N-2D}
\end{figure}
\begin{figure}[h]
\centering
\includegraphics[scale=0.5]{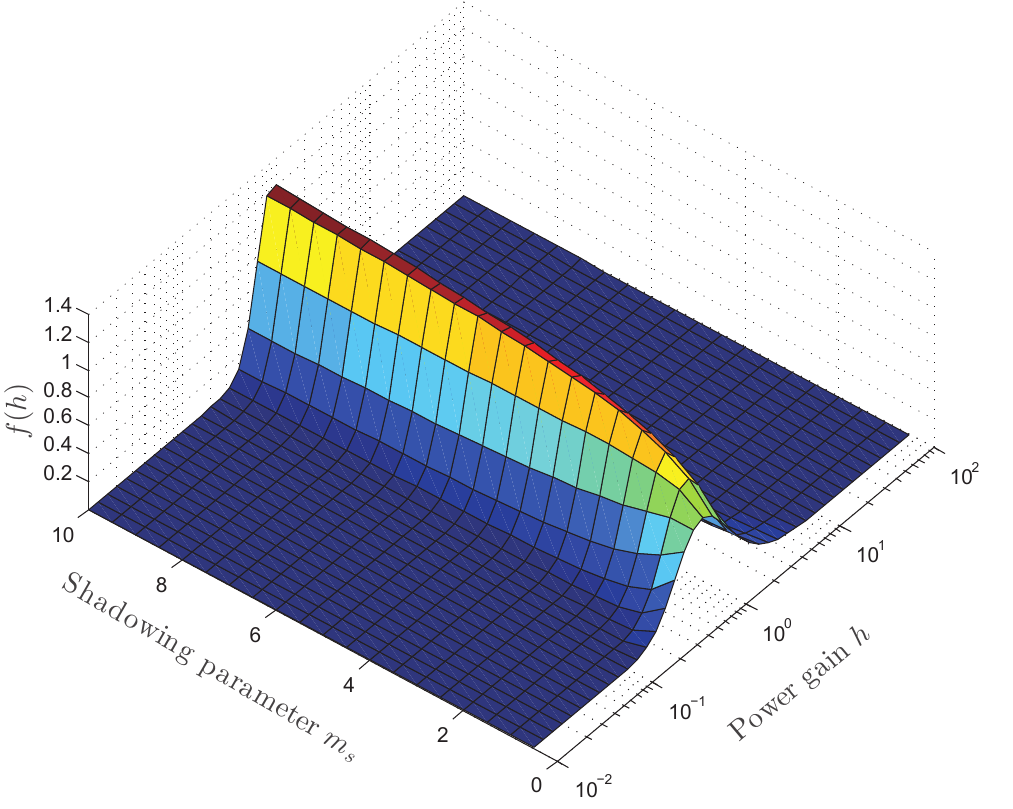}
\caption{The PDF of heterogeneous $\mathcal{F}$ composite fading channel versus the channel characteristics of power gain $h$ and shadowing parameter $m_{s}$.} \label{figure-pdf-ms_h}
\end{figure}
\begin{figure}[h]
\centering
\includegraphics[scale=0.5]{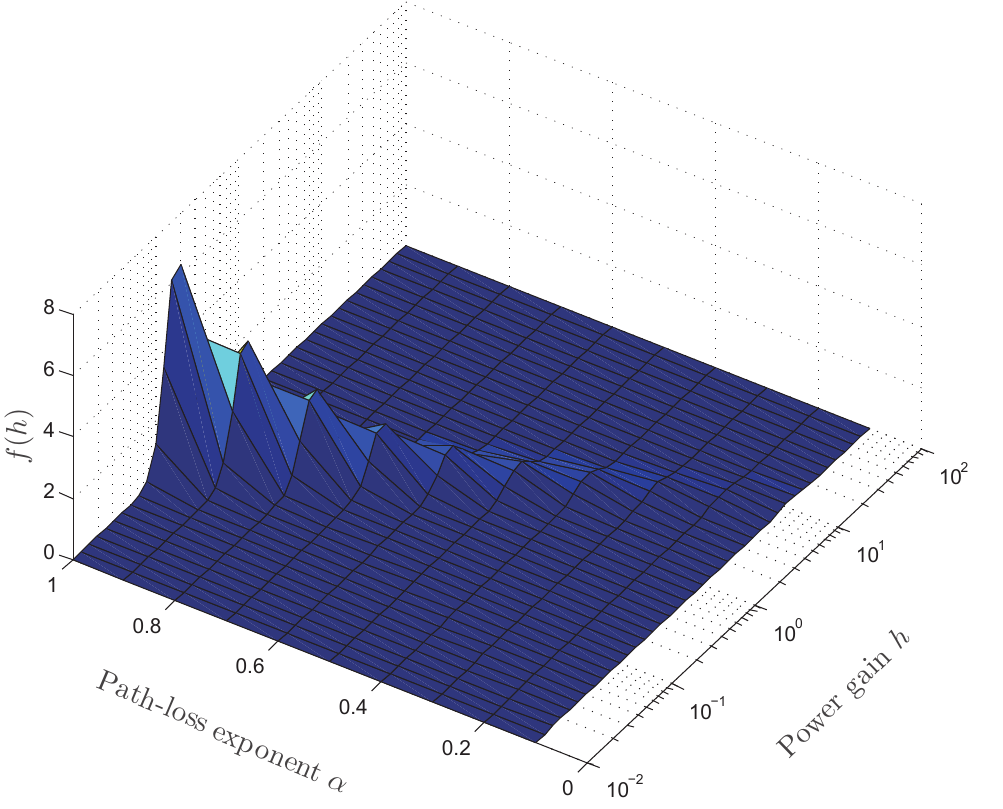}
\caption{The PDF of heterogeneous $\mathcal{F}$ composite fading channel versus the channel characteristics of power gain $h$ and the path-loss exponent $\alpha$.} \label{figure-pdf-alpha_h}
\end{figure}
To better understand the insights of the PDF of heterogeneous $\mathcal{F}$ composite fading channel, we depict Figs.~\ref{figure-pdf-K_h},~\ref{figure-pdf-N_h},~\ref{pdf-N-2D},~\ref{figure-pdf-ms_h}, and~\ref{figure-pdf-alpha_h} with typical values for channel parameters. Fig.~\ref{figure-pdf-K_h} shows the PDF of heterogeneous $\mathcal{F}$ composite fading channel versus the channel characteristics of power gain $h$ and the number of 5G-UAVs $K$. As shown in Fig.~\ref{figure-pdf-K_h}, when the number of 5G-UAVs increases, the PDF of heterogeneous $\mathcal{F}$ composite fading channel shifts to the right, which tends towards a large power gain. Therefore, we find that the average power gain $h$ increases as the number of 5G-UAVs $K$ increases.

Figure~\ref{figure-pdf-N_h} shows the PDF of heterogeneous $\mathcal{F}$ composite fading channel versus the channel characteristics of power gain $h$ and the number of reflectors $N$. For the numerical analyses, we divide Eq.~\eqref{eq:_fh} into three parts as $X_{1}=\frac{h^{KNm-1}\prod_{k=1}^{K}L_{k}^{\alpha}}{B\left(KNm, KNm_{s}\right)}$, $X_{2}=(\frac{\Lambda}{K})^{KNm}$, and $X_{3}={}_{2} F_{1}\left( KN(m+m_{s}), KNm ; KNm; -\frac{\Lambda h}{K}\right)$, which all increase at first and then decrease as $N$ increases. To further clarify the characteristics, we plot the PDF of heterogeneous $\mathcal{F}$ composite fading channel versus the power gain $h$ with different numbers of reflectors $N$, as shown in Fig.~\ref{pdf-N-2D}. It is clear that the curve shifts to the right corresponding to large power gains as $N$ increases. Therefore, the PDF of heterogeneous $\mathcal{F}$ composite fading channel, which can be rewritten as $f(h)=X_{1}X_{2}X_{3}$, is drawn as Fig.~\ref{figure-pdf-N_h}. Numerical results show that the power gain increases as the number of reflectors $N$ increases~\cite{8647620}~\cite{8746155}.

Figures.~\ref{figure-pdf-ms_h} and~\ref{figure-pdf-alpha_h} show the minor changes that the power gain very likely increases as $m_{s}$ increases and $\alpha$ decreases. In detail, Fig.~\ref{figure-pdf-ms_h} shows the PDF of heterogeneous $\mathcal{F}$ composite fading channel versus the channel characteristics of power gain $h$ and shadowing parameter $m_{s}$. It can be seen from Fig.~\ref{figure-pdf-ms_h} that the range of achievable power gain is gradually concentrated as the shadowing parameter increases. This is because that the links undergo more slightly shadowing as $m_{s}$ increases. Then, Fig.~\ref{figure-pdf-alpha_h} shows the PDF of heterogeneous $\mathcal{F}$ composite fading channel versus the channel characteristics of power gain $h$ and the path-loss exponent $\alpha$. We observe that the average power gain also increases as the path-loss exponent $\alpha$ decreases. In summary, there is a better average power gain as the number of reflectors $N$ increases, the path-loss exponent $\alpha$ decreases, and the shadowing parameter $m_{s}$ increases, which are consistent with the practical emergency communication scenario.

\subsection{The Heterogeneous $\mathcal{F}$ Composite Fading Channel Adapted Optimal Power Allocation Scheme}
Due to the limitation of available bandwidth and power resources in the poster-disaster areas, highly efficient resource allocation schemes are very important for EWC networks. In the following, we develop the optimal power allocation scheme with the simple closed-form expression.

The ergodic capacity corresponding to the links between the on-site command center and the trapped user, denoted by $C_{1}$, can be written as follows:
\begin{eqnarray}
C_{1}= \int_{0}^{\infty}B\log_{2}\left[1+\frac{P(h)}{\delta^{2}}h\right]f(h)\mathrm{d}h,
\label{eq:C1}
\end{eqnarray}
where $P(h)$ is the instantaneous transmit power and all the bandwidth is utilized. Then, we formulate the ergodic capacity maximization problem, denoted by ${\bf \emph{P}1}$, for 5G-UAVs based EWC networks, as follows:
\begin{eqnarray}
\label{problem_maxC1}
{\bf \emph{P}1:}\arg\max\limits _{P(h)} \left\{ \int_{0}^{\infty}B\log_{2}\left[1+\frac{P(h)}{\delta^{2}}h\right]f(h)\mathrm{d}h \right\}
\end{eqnarray}
\begin{eqnarray}
\text{s.t.}
\left\{\begin{array}{ll}
\vspace{0.2cm}\int_{0}^{\infty}P(h)f(h)\mathrm{d}h\leq \bar{P}_{h}; \\
0\leq P(h)\leq P_{h},\\
\end{array}\right. \nonumber
\end{eqnarray}
where $\bar{P}_{h}$ is the average transmit power and $P_{h}$ is the peak power constraint. For the objective function of problem {\bf \emph{P}1} specified in Eq.~\eqref{problem_maxC1}, we denote by $G_{1}(P)=B\log_{2}\left[1+\frac{P(h)}{\delta^{2}}h\right]f(h)$.  Since $\frac{\partial^{2} G_{1}(P)}{\partial P^{2}(h)}=-\left(\frac{h/\delta^{2}}{1+hP(h)/\delta^{2}}\right)^{2}Bf(h)/\ln2 \leq 0$ holds, the objective function of problem ${\bf \emph{P}1}$ is concave with respect to the $P(h)$. Also, the items $P(h)f(h)$ and $P(h)$ are both linear with respect to $P(h)$, respectively. Thus, the problem ${\bf \emph{P}1}$ is a strictly convex optimization problem.

To solve problem ${\bf \emph{P}1}$, we formulate the Lagrangian function, denoted by $J_{1}$, as follows:
\begin{eqnarray}
\begin{aligned}
&J_{1}\!=\!\int_{0}^{\infty}B\log_{2}\left[1\!+\!\frac{P(h)}{\delta^{2}}h\right]f(h)\mathrm{d}h\!-\!\lambda_{0}\left[P(h)\!- \!P_{h}\right]\\
&\hspace{3cm} -\lambda\left[\int_{0}^{\infty}P(h)f(h)\mathrm{d}h-\bar{P}_{h}\right],
\end{aligned}
\end{eqnarray}
where $\lambda$ and $\lambda_{0}$ are the Lagrangian multipliers corresponding to the average transmit power constraint and the peak transmit power constraint, respectively. Then, the Karush-Kuhn-Tucker (KKT) conditions for problem {\bf \emph{P}1} can be derived as follows:
\begin{eqnarray}
\begin{aligned}
\left\{\begin{array}{l}
\vspace{0.2cm}\left[\left(\dfrac{B/\ln2}{1+hP(h)/\delta^{2}}\right)\dfrac{h}{\delta^{2}}-\lambda\right]f(h)-\lambda_{0}=0; \\
\vspace{0.2cm}\lambda\left[\int_{0}^{\infty}P(h)f(h)\mathrm{d}h-\bar{P}_{h}\right]=0; \\
\vspace{0.2cm}\lambda_{0}\left[P(h)- P_{h}\right]=0;\\
\vspace{0.2cm}P_{h}, P(h), \lambda, \lambda_{0} \geq 0.
\end{array}\right.
\end{aligned}\label{eqs:st_p1}
\end{eqnarray}
Then, the optimal power allocation scheme for problem {\bf \emph{P}1}, denoted by $P(h)$, can be derived as follows:
\begin{equation}
\label{eq:opt}
\begin{aligned}
P(h)=\left\{\begin{array}{ll}
\vspace{0.2cm}\dfrac{\delta^{2}}{h_{0}}-\dfrac{\delta^{2}}{h}, & \text { if } \quad h \geqslant h_{0}, P(h)<P_{h}; \\
\vspace{0.2cm}P_{h}, & \text { if } \quad h \geqslant h_{0}, P(h) \geq P_{h}; \\
\vspace{0.2cm}0, & \text { if } \quad h<h_{0},
\end{array}\right.
\end{aligned}
\end{equation}
where $h_{0}$ is the outage threshold.
According to Eqs.~\eqref{eqs:st_p1} and~\eqref{eq:opt}, as $\lambda\neq 0$, we have
\begin{equation}
\begin{aligned}
\int_{h_{0}}^{\infty}\left(\frac{\delta^{2}}{h_{0}}-\frac{\delta^{2}}{h}\right) f(h) \mathrm{d} h=\bar{P}_{h},
\end{aligned}\label{eq:h0-inf}
\end{equation}
which can be rewritten as follows:
\begin{equation}
\begin{aligned}
\int_{0}^{\infty}\!\!\left(\frac{1}{h_{0}}\!-\!\frac{1}{h}\right) f(h) \mathrm{d} h\!\!-\!\!\int_{0}^{h_{0}}\!\!\left(\frac{1}{h_{0}}\!-\!\frac{1}{h}\right) f(h) \mathrm{d} h\!=\!\frac{\bar{P}_{h}}{\delta^{2}}.
\end{aligned}
\end{equation}
Using~\cite[Eq.~(8)]{8359199}, we can obtain
\begin{equation}
\begin{aligned}
&\int_{0}^{h_{0}} \frac{1}{h} f(h) \mathrm{d} h =\int_{0}^{h_{0}} \frac{1}{h_{0}} f(h) \mathrm{d} h \\
&\hspace{0.2cm}=\frac{\prod\limits_{k=1}^{K}L_{k}^{\alpha} \Gamma\left(K N m+K N m_{s}\right)}{h_{0} \Gamma(1+K N m) \Gamma\left(K N m_{s}\right)}\left(\frac{\Lambda h_{0}}{K N}\right)^{K N m} \\
&\hspace{0.2cm} \times{ }_{2} F_{1}\left(K N\left(m+m_{s}\right), K N m ; 1+K N m ; \frac{-\Lambda h_{0}}{K N}\right).
\end{aligned}
\end{equation}
Then, Eq.~\eqref{eq:h0-inf} can be rewritten as follows:
\begin{equation}
\begin{aligned}
\int_{0}^{\infty}\left(\frac{\delta^{2}}{h_{0}}-\frac{\delta^{2}}{h}\right) f(h) \mathrm{d} h=\bar{P}_{h}.
\end{aligned}\label{eq:0-inf}
\end{equation}
According to~\cite[Eq.~(2.24.2.1)]{book} given by
\begin{equation}
\begin{aligned}
&\mathlarger{\int}_{0}^{\infty} \tau^{\varsigma-1} G_{u v}^{s t}\left(\xi \tau \bigg| \left.\begin{array}{l}
\left(a_{u}\right) \\
\left(b_{v}\right)
\end{array}\right.\right) d \tau \\
&\hspace{0.2cm}\vspace{0.2cm}=\xi^{-\varsigma} \Gamma\left[\begin{array}{l}
\left(b_{s}\right)+\varsigma, 1-\left(a_{t}\right)-\varsigma \\
a_{t+1}\!+\!\varsigma, \ldots, a_{u}\!\!+\!\!\varsigma, 1\!\!-\!\!b_{s+1}\!\!-\!\!\varsigma, \ldots, 1\!\!-\!\!b_{v}\!\!-\!\!\varsigma
\end{array}\right],
\end{aligned}\label{non-G}
\end{equation}
where $\tau$, $\varsigma$, $\xi$, $s$, $t$, $u$, $v$, $a_{u}$, $b_{v}$, $b_{s}$, $a_{t}$, $a_{t+1}$, and $b_{s+1}$ are corresponding parameters,
we can obtain
\begin{equation}
\begin{aligned}
\left\{\begin{array}{l}
\vspace{0.2cm}\!\!\mathlarger{\int}_{0}^{\infty}\!\!\dfrac{1}{h_{0}}f(h)\mathrm{d} h= \dfrac{1}{h_{0}}\prod\limits_{k=1}^{K}L_{k}^{\alpha};\\
\vspace{0.2cm}\!\!\mathlarger{\int}_{0}^{\infty}\!\!\dfrac{1}{h}f(h)\mathrm{d} h\!=\! \dfrac{\Gamma(K N m\!\!-\!\!1) \Gamma\left(K N m_{s}\!\!+\!\!1\right)}{\Gamma(K N m) \Gamma\left(K N m_{s}\right)}\! \times \! \dfrac{\Lambda}{K}\!\!\prod\limits_{k=1}^{K}L_{k}^{\alpha}.
\end{array}\right.
\end{aligned}\label{eq:derivation}
\end{equation}
Substituting Eq.~\eqref{eq:derivation} into Eq.~\eqref{eq:0-inf}, we have
\begin{equation}
\label{eq:h0}
\begin{aligned}
h_{0}\!\!=\!\!\prod\limits_{k=1}^{K}L_{k}^{\alpha}\!\!\left[\frac{\bar{P}_{h}}{\delta^{2}}\!\!+\!\!\dfrac{\Gamma(K N m\!\!-\!\!1) \Gamma\left(K N m_{s}\!\!+\!\!1\right)}{\Gamma(K N m) \Gamma\left(K N m_{s}\right)}\frac{\Lambda}{K} \!\!\prod\limits_{k=1}^{K}L_{k}^{\alpha}\right]^{\!-\!1}.
\end{aligned}
\end{equation}
Then, substituting Eq.~\eqref{eq:h0} into Eq.~\eqref{eq:opt}, the optimal power allocation scheme for problem {\bf \emph{P}1} is obtained as Eq.~\eqref{eq:power_C1}.
\begin{figure*}
\begin{equation}
\label{eq:power_C1}
\begin{split}
P(h)=
\left\{\begin{array}{ll}
\vspace{0.2cm}\left[\bar{P}_{h}+\dfrac{\delta^{2}\Lambda\Gamma(KNm\!-\!1)\Gamma(KNm_{s}+1)}{K\Gamma (KNm)\Gamma (KNm_{s})}\prod\limits_{k=1}^{K}L_{k}^{\alpha}\right] \left(\prod\limits_{k=1}^{K}L_{k}^{\alpha}\right)^{-1}-\dfrac{\delta^{2}}{h}, &\text{if}\quad h \geqslant h_{0}, P(h)<P_{h}; \\
\vspace{0.2cm} P_{h}, &\text{if}\quad h \geqslant h_{0}, P(h)\geq P_{h};\\
\vspace{0.2cm} 0, &\text{if}\quad h < h_{0}.
\end{array}\right.
\end{split}
\end{equation}
\hrulefill
\end{figure*}
\begin{figure*}
\begin{equation}
\label{eq:power_C1_A}
\begin{split}
\widetilde{P}(h)=
\left\{\begin{array}{ll}
\vspace{0.2cm}\left(\bar{P}_{h}+\dfrac{\epsilon \Lambda \delta^{2}}{K}\prod\limits_{k=1}^{K}L_{k}^{\alpha}\right)\left(\prod\limits_{k=1}^{K}L_{k}^{\alpha}\right)^{-1}-\dfrac{\delta^{2}}{h}, &\text{if}\quad h_{0}\leq h < \left(\delta^{2}\prod\limits_{k=1}^{K}L_{k}^{\alpha}\right)\left[\bar{P}_{h}+(\frac{\epsilon \Lambda \delta^{2}}{K}-P_{h})\prod\limits_{k=1}^{K}L_{k}^{\alpha}\right]^{-1}; \\
\vspace{0.2cm} P_{h}, &\text{if}\quad h \geqslant \left(\delta^{2}\prod\limits_{k=1}^{K}L_{k}^{\alpha}\right)\left[\bar{P}_{h}+(\frac{\epsilon \Lambda \delta^{2}}{K}-P_{h})\prod\limits_{k=1}^{K}L_{k}^{\alpha}\right]^{-1};\\
\vspace{0.2cm} 0, &\text{if}\quad h < h_{0},
\end{array}\right.
\end{split}
\end{equation}
\hrulefill
\end{figure*}
\begin{figure*}
\begin{equation}
\label{eq:opt_C1}
\begin{split}
C_{1} =\displaystyle{\displaystyle\int_{h_{0}}^{\infty}{B \log_{2}\left\{1+\text{min}\left\{\left[\dfrac{\bar{P}_{h}+\frac{\epsilon m \bar{L}^{\alpha} \delta^{2}}{KNm_{s}\bar{h}}\prod\limits_{k=1}^{K}L_{k}^{\alpha}}{\prod\limits_{k=1}^{K}L_{k}^{\alpha}}-\dfrac{\delta^{2}}{h}\right]^{+},~P_{h}\right\}
\dfrac{h}{\delta^{2}}\right\}f(h)\mathrm{d}h}}.
\end{split}
\end{equation}
\hrulefill
\end{figure*}

According to literatures~\cite{8474223}~\cite{7886273}, the Fisher-Snedecor $\mathcal{F}$ model includes several fading distributions as special cases, such as Nakagami-$m$ ($m_{s}\rightarrow \infty$) and Rayleigh ($m_{s}\rightarrow \infty$, $m=1$) distributions. As $m_{s}\rightarrow 0$, the channel undergoes heavier shadowing. In the typical emergence communication scenarios, we consider the general cases which are mainly with heavy shadowing. Thus, the asymptotic expressions of the transmit power $\widetilde{P}(h)$ is derived as Eq.~\eqref{eq:power_C1_A} using the approximation for $\frac{\Gamma(KNm-1)\Gamma(KNm_{s}+1)}{\Gamma (KNm)\Gamma (KNm_{s})}\simeq \epsilon$, where $m_{s}\leq m$ and $0<\epsilon \leq1$.

\begin{figure}[h]
\centering
\vspace{-2pt}
\includegraphics[scale=0.50]{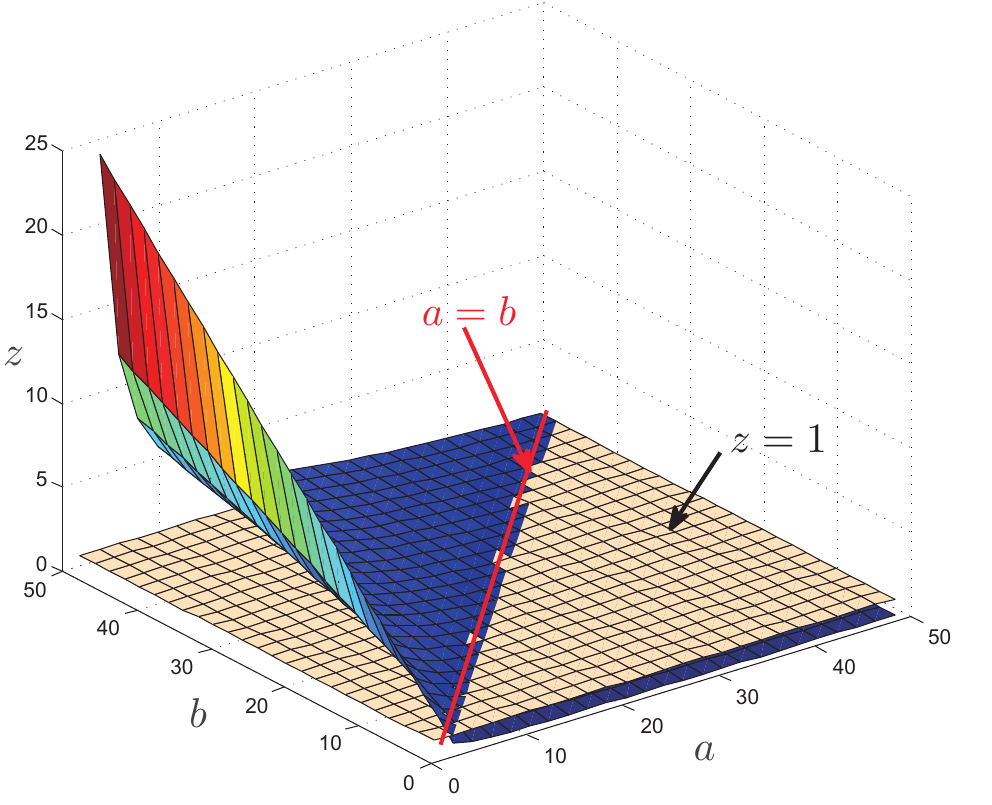}
\caption{The term $z=\frac{\Gamma(a-1)\Gamma(b+1)}{\Gamma (a)\Gamma (b)}$ with $a$ and $b$.} \label{gamma_proof}
\vspace{-8pt}
\end{figure}

Here, we show the figure for $\frac{\Gamma(a-1)\Gamma(b+1)}{\Gamma (a)\Gamma (b)}\simeq \epsilon$, where $0<\epsilon \leq1, a\geq b$. As shown in Fig.~\ref{gamma_proof}, when $a\geq b$ holds,  $z$ is approximately a plane, which is lower than the plane $z=1$.

\begin{figure}[h]
\centering
\vspace{-2pt}
\includegraphics[scale=0.5]{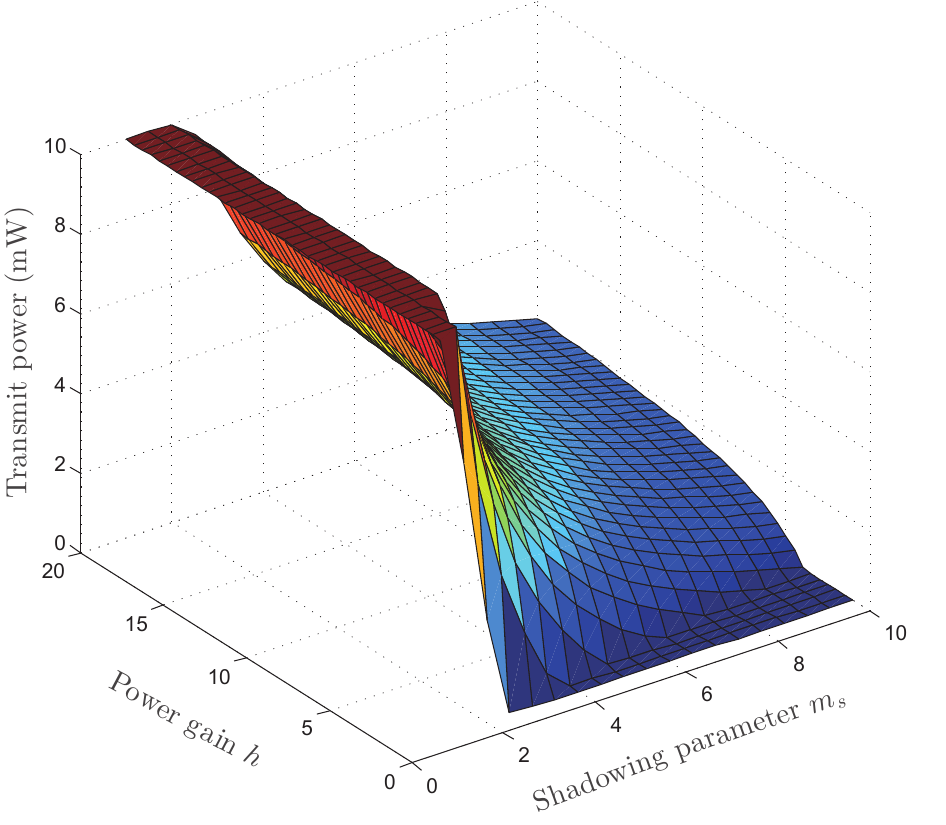}
\caption{The heterogeneous $\mathcal{F}$ composite fading channel adapted optimal power allocation scheme versus the power gain $h$ and the shadowing parameter $m_{s}$ of heterogeneous $\mathcal{F}$ composite channel.} \label{power_allocation_C1}
\vspace{-8pt}
\end{figure}

\begin{figure}[h]
\centering
\vspace{-2pt}
\includegraphics[scale=0.5]{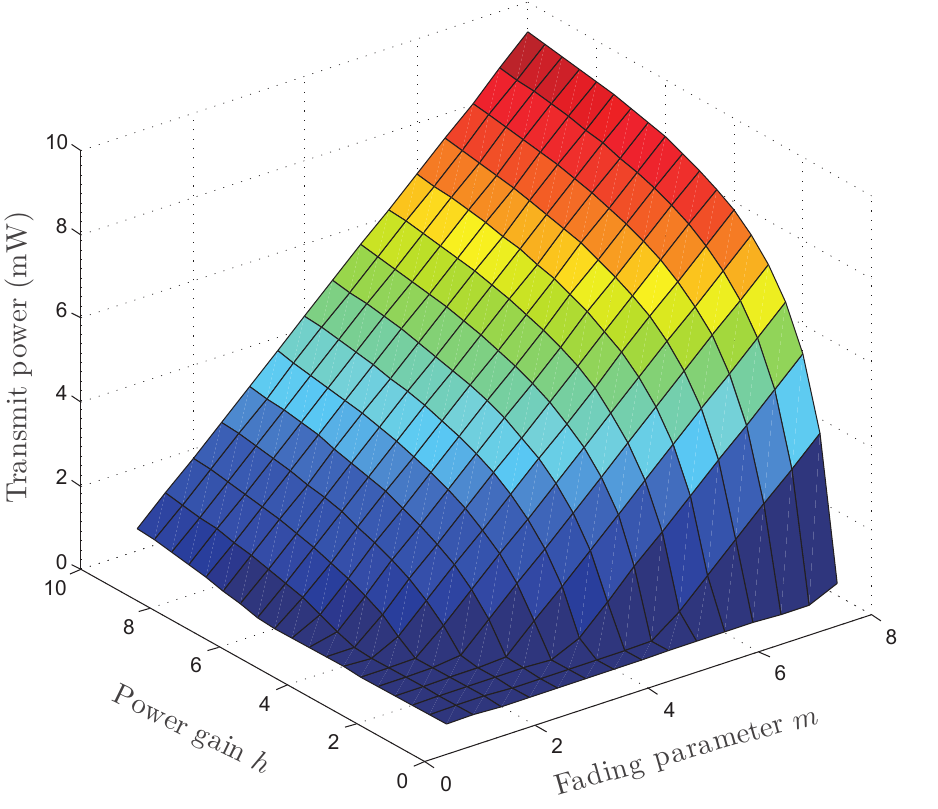}
\caption{The heterogeneous $\mathcal{F}$ composite fading channel adapted optimal power allocation scheme versus the power gain $h$  and the fading parameter $m$ of heterogeneous $\mathcal{F}$ composite channel.} \label{p_h_m}
\vspace{-8pt}
\end{figure}

\begin{figure}[h]
\centering
\vspace{-2pt}
\includegraphics[scale=0.5]{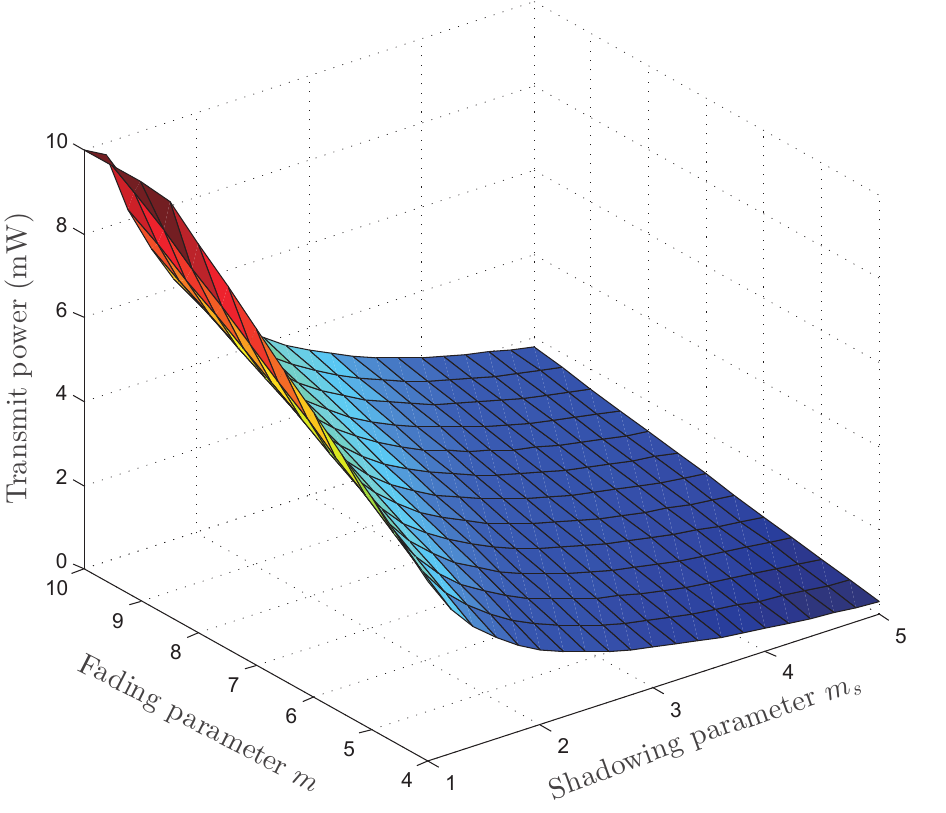}
\caption{The heterogeneous $\mathcal{F}$ composite fading channel adapted optimal power allocation scheme with the shadowing parameter $m_{s}$ and the fading parameter $m$.} \label{p_ms_m}
\vspace{-8pt}
\end{figure}

With the power allocation scheme, the maximum channel capacity is derived as Eq.~\eqref{eq:opt_C1}, where $[\xi]^{+}\triangleq \text{max}\{\xi,0\}$. To better understand our proposed the optimal power allocation scheme, we plot the instantaneous power under the heterogeneous $\mathcal{F}$ composite fading channel in Figs.~\ref{power_allocation_C1}, ~\ref{p_h_m}, and~\ref{p_ms_m}, respectively. As illustrated in Fig.~\ref{power_allocation_C1}, we show the optimal power allocation scheme versus the power gain $h$ and the shadowing parameter $m_{s}$ of the heterogeneous $\mathcal{F}$ composite channel. It can be seen from the three-dimensional graph that as the power gain increases, the corresponding channel is allocated more power. On the other hand, more power is allocated to the channel as the shadowing parameter $m_{s}$ decreases. Since the signal undergoes heavier shadowing as $m_{s}$ decreases, more power is needed to satisfy the basic communication requirements for emergency communications.
Figure~\ref{p_h_m} shows the optimal power allocation scheme with the power gain $h$ and the fading parameter $m$. We can find that the allocated transmit power increases as both the power gain and fading parameter increase, which corresponding to better channel condition, thus achieving higher transmission rates.

Similarly, it is clear to find that the relationships among the transmit power $P(h)$, fading parameter $m$, and the shadowing parameter $m_{s}$ from Fig.~\ref{p_ms_m}. It can be seen from Fig.~\ref{p_ms_m} that the allocated power increases as the fading parameter $m$ increases while the shadowing parameter $m_{s}$ decreases. This is because the shadowing parameter slightly impacts the power gain as shown in Fig.~\ref{figure-pdf-ms_h}. Then, the fading parameter $m$ dominates the optimal power allocation under this case for achieving the higher transmission rates at the better power gain.


\subsection{Joint Bandwidth-Power Allocation Scheme}

In typical emergency communication scenarios, the channel states dynamically change due to the destruction of the natural environments such as secondary disasters. Even when multiple 5G-UAVs cooperate, the received SNR is relatively low. Therefore, we further refine the resource allocation strategy. The available bandwidth and power resources are allocated to $K$ UAVs with adaptation to the instantaneous channel state for EWC scenarios. Under such a case, the ergodic capacity corresponding to the links between the on-site command center and the trapped user, denoted by $C_{2}(B_{k}, P_{2}(h_{k}))$, can be rewritten as follows:
\begin{align}
C_{2}(B_{k}, P_{2}(h_{k}))\!=\!\underbrace{\iint\ldots\int_{0}^{\infty}}_{K-fold}\sum\limits_{k=1}^{K}B_{k}\log_{2}\left[1+\frac{P_{2}(h_{k})h_{k}}{N_{0}B_{k}}\right]\nonumber\\
&\hspace{-5cm}\times\prod\limits_{k=1}^{K}f_{k}(h_{k})\mathrm{d}h_{1}\ldots\mathrm{d}h_{K},
\label{eq:C2}
\end{align}
where $B_{k}$ and $P_{2}(h_{k})$ denote the allocated bandwidth and transmit power of the $k$-th sub-channel, respectively.

Then, we formulate the ergodic capacity maximization problem, denoted by ${\bf \emph{P}2}$, for multiple 5G-UAVs based EWC networks, as follows:
\begin{eqnarray}
\label{problem_maxC2}
{\bf \emph{P}2:}\arg\max\limits _{B_{k}, P_{2}(h_{k})}\left\{ \sum\limits_{k=1}^{K}B_{k}\log_{2}\left[1+\frac{P_{2}(h_{k})h_{k}}{N_{0}B_{k}}\right]\right\}
\end{eqnarray}
\begin{eqnarray}
\text{s.t.}
\left\{\begin{array}{ll}
\vspace{0.2cm}P_{2}(h_{k})\geq 0, \quad k=1,2,3,...,K; \\
\vspace{0.2cm}B_{k} \geq 0, \quad k=1,2,3,...,K; \\
\vspace{0.2cm}\sum_{k=1}^{K}P_{2}(h_{k}) \leq P_{h}, \quad k=1,2,3,...,K; \\
\vspace{0.2cm}\sum_{k=1}^{K}B_{k} \leq B, \quad k=1,2,3,...,K.
\end{array}\right.\nonumber
\end{eqnarray}
For the objective function of problem ${\bf \emph{P}2}$ specified in Eq.~\eqref{problem_maxC2}, we denote by $G_{2}\left(B_{k}, P_{2}(h_{k})\right)=B_{k}\log_{2}\left[1+\dfrac{P_{2}(h_{k})h_{k}}{N_{0}B_{k}}\right]$.
The Hessian Matrix of $G_{2}\left(B_{k}, P_{2}(h_{k})\right)$, denoted by $H(G_{2})$, is derived as follows:
\begin{equation}
\begin{aligned}
&H\left(G_{2}\right)=\left[\begin{array}{c}
\vspace{0.3cm}\dfrac{\partial^{2} G_{2}}{\partial B_{k}^{2}} \qquad \dfrac{\partial^{2} G_{2}}{\partial B_{k} \partial P_{2}\left(h_{k}\right)} \\
\vspace{0.3cm}\dfrac{\partial^{2} G_{2}}{\partial P_{2} \left(h_{k}\right) \partial B_{k}} \qquad \dfrac{\partial^{2} G_{2}}{\partial P_{2}^{2}\left(h_{k}\right)}
\end{array}\right] \\
&\vspace{0.2cm}=\left[\!\!\begin{array}{c}
\vspace{0.3cm}\dfrac{-\left(P_{2} (h_{k}) h_{k}\right)^{2}/\ln2}{B\left(N_{0} B_{k}+P_{2} (h_{k})h_{k}\right)^{2}} \quad \dfrac{P_{2}(h_{k}) h_{k}^{2}/\ln2}{\left(N_{0} B_{k}+P_{2}(h_{k}) h_{k}\right)^{2}} \\
\dfrac{P_{2}(h_{k}) h_{k}^{2}/\ln2}{\left(N_{0} B_{k}+P_{2}(h_{k}) h_{k}\right)^{2}} \quad \dfrac{-\left(B_{k} h_{k}\right)^{2}/\ln2}{\left(N_{0} B_{k}+P_{2} (h_{k})h_{k}\right)^{2} }
\end{array}\!\!\right].
\end{aligned}
\end{equation}
Since $|H\left(G_{2}\right)|=0$ holds, the objective function of problem ${\bf \emph{P}2}$ is concave with respect to $\{B_{k}, P_{2}(h_{k})\}$. Also, the items $P_{2}(h_{k})$ and $B_{k}$ are both linear with respect to $\{B_{k}, P_{2}(h_{k})\}$, respectively. Therefore, the problem ${\bf \emph{P}2}$ is a strictly convex optimization problem.

%
%

To solve problem ${\bf \emph{P}2}$, we formulate the Lagrangian function, denoted by $J_{2}$, as follows:
\begin{eqnarray}
\begin{aligned}
J_{2}&=\sum\limits_{k=1}^{K}B_{k}\log_{2}\left[1+\frac{P_{2}(h_{k})}{N_{0}B_{k}}h_{k}\right]+\sum\limits_{k=1}^{K}\mu_{k}B_{k}\\
&\hspace{1cm}+\sum\limits_{k=1}^{K}\lambda_{k}P_{2}(h_{k})-\chi\left(\sum_{k=1}^{K}B_{k}-B\right)\\
&\quad\quad\quad\quad\quad\quad\quad\quad -\omega\left[\sum_{k=1}^{K} P_{2}(h_{k}) - P_{h}\right],\\
\end{aligned}
\end{eqnarray}
where $\lambda_{k} (1\leq k \leq 1)$ is the Lagrangian multipliers corresponding to the constraints that force the power allocated to each sub-channel, $\mu_{k} (1\leq k \leq 1)$ is the Lagrangian multipliers corresponding to the constraints that force the bandwidth allocated to each channel, $\omega$ denotes the Lagrangian multiplier corresponding to the total power constraint, and $\chi$ is the Lagrangian multiplier corresponding to the total of available bandwidth constraint. Then, the KKT conditions for problem~${\bf \emph{P}2}$ can be derived as follows:

\begin{align}
\left\{\begin{array}{ll}
\dfrac{\partial J_{2}}{\partial P_{2}(h_{k})}\!=\!\dfrac{B_{k}}{\ln 2}\dfrac{h_{k}}{\delta^{2}+P_{2}(h_{k})h_{k}}\!+\lambda_{k}\!-\!\omega=0,\\
\quad\quad\quad\quad\quad\quad\quad\quad\quad\quad\quad\quad\quad\quad k=1,2,3,...,K;\\
\dfrac{\partial J_{2}}{\partial B_{k}}\!\!=\!\!\log_{2}\left[\!1\!+\!\dfrac{P_{2}(h_{k})h_{k}}{N_{0}B_{k}}\!\right]\!\!-\!\!\dfrac{P_{2}(h_{k})h_{k}/\ln2}{N_{0}B_{k}\!\!+\!\!P_{2}(h_{k})h_{k}}\!\!+\!\!\mu_{k}\!\!-\!\!\chi=0,\\
\quad\quad\quad\quad\quad\quad\quad\quad\quad\quad\quad\quad\quad\quad k=1,2,3,...,K;\\
\vspace{0.2cm}\lambda_{k}P_{2}(h_{k})=0, \quad k=1,2,3,...,K;\\
\vspace{0.2cm}\mu_{k}B_{k}=0, \quad k=1,2,3,...,K;\\
\vspace{0.2cm}\omega\left[\sum_{k=1}^{K}P_{2}(h_{k}) - P_{h}\right]=0;\\
\vspace{0.2cm}\chi\left(\sum_{k=1}^{K}B_{k}-B\right)=0;\\
\vspace{0.2cm}\lambda_{k},~\mu_{k} \geq 0, \quad k=1,2,3,...,K;\\
\vspace{0.2cm}P_{2}(h_{k}),~B_{k} \geq 0, \quad k=1,2,3,...,K;\\
\vspace{0.2cm}P_{h},~B,~\omega,~\chi \geq 0.\\
\end{array}\right.
\label{eq:C2-kkt}
\end{align}

It is hard to directly solve Eq.~\eqref{eq:C2-kkt} to obtain the closed-form solution of problem ${\bf \emph{P}2}$. First, we derive the sub-optimal solution with given bandwidth or power limitation under the KKT conditions. Then, we use the gradient descent method to obtain the optimal solution along the gradient direction based on the sub-optimal solutions. These two steps are illustrated in {\bf Algorithm 1} and can achieve the optimal solutions for problem {\bf \emph{P}2}~\cite{6168145},
where $\beta$ is the step factor, $\triangle$ is the stopping threshold, $i_{\rm{max}}$ is the maximum number of iterations, and ${\bf \nabla J_{2}}$ is the direction gradient. Note that $\Upsilon^{(i)}$ is the $i$-th iteration of $\Upsilon$, where $\Upsilon$ can be $B_{k}, P_{2}(h_{k}), \lambda_{k}, \mu_{k}, \omega, \chi$, and $C_{2}$. Under {\bf Algorithm 1}, the ergodic capacity is updated with $B_{k}^{(i+1)}$ and $P_{2}(h_{k})^{(i+1)}$ after each iteration. The iteration factors $B_{k}^{(i+1)}$,  $P_{2}(h_{k})^{(i+1)}$, $\lambda_{k}^{(i+1)}$, $\mu_{k}^{(i+1)}$, $\omega^{(i+1)}$, and $\chi^{(i+1)}$ can be iteratively derived until satisfying the convergence condition, i.e., $C_{2}^{(i+1)}-C_{2}^{(i)}\geq \Delta$ or $i\leq i_{\rm{max}}$. When the increment of capacity is less than a very small value or the number of iterations is larger than the given threshold, the iteration is terminated. Thus, the convergence of {\bf Algorithm 1} is guaranteed. Then, the optimal resource allocation solution $[B_{k}^{*}, P_{2}(h_{k})^{*}]$ can be found to maximize the ergodic capacity.
\begin{figure*}
\begin{equation}
\label{eq:Bk_i+1}
\begin{split}
\left\{\begin{array}{ll}
\vspace{0.2cm}B_{k}^{(i+1)}=\left\{B_{k}^{(i)}+\beta \left[\log_{2}(1+\dfrac{P_{2}(h_{k})^{(i)}h_{k}}{N_{0}B_{k}^{(i)}}) -\dfrac{P_{2}(h_{k})^{(i)}h_{k}/\ln2}{N_{0}B_{k}^{(i)}+P_{2}(h_{k})^{(i)}h_{k}}+\mu_{k}^{(i)}-\chi^{(i)}\right]\right\}^{+}, \forall k;\\
\vspace{0.2cm}P_{2}(h_{k})^{(i+1)}=B_{k}^{(i+1)}\left[\dfrac{1}{(-\lambda_{k}^{(i)}+\omega^{(i)})\ln2}-\dfrac{N_{0}}{h_{k}}\right]^{+}, \forall k;\\
\vspace{0.2cm}\chi^{(i+1)}=\left[\chi^{(i)}+\beta \left(\sum_{k=1}^{K}B_{k}^{(i+1)}-B\right)\right]^{+}, \forall k; \quad \mu_{k}^{(i+1)}=\left[\mu_{k}^{(i)}+\beta B_{k}^{(i+1)}\right]^{+}, \forall k;\\
\vspace{0.2cm}\omega^{(i+1)}=\left[\omega^{(i)}+\beta \left(\sum_{k=1}^{K}P_{2}(h_{k})^{(i+1)}-P_{h}\right)\right]^{+}, \forall k; \quad \lambda_{k}^{(i+1)}=\left[\lambda_{k}^{(i)}+\beta P_{2}(h_{k})^{(i+1)}\right]^{+}, \forall k.\\
\end{array}\right.
\end{split}
\end{equation}
\hrulefill
\end{figure*}

\begin{algorithm}[h]
\label{Code:1}
\caption{Joint Bandwidth-Power Allocation Algorithm}
\begin{algorithmic}[1]
\STATE Initialize $\beta$, $\triangle$, $i_{\rm{max}}$ and ${\bf \nabla J_{2}}$;
\STATE \vspace{0.1cm}Initialize the source transmit power peak $P_{h}$ and the total available bandwidth $B$;
\STATE \vspace{0.1cm}{\bf If} ($i=0$) {\bf then}
\STATE \vspace{0.1cm}\hspace{0.35cm} Initialize $B_{k}^{(0)}$ and $P_{2}(h_{k})^{(0)}$, $\forall k$;
\STATE \vspace{0.1cm}\hspace{0.35cm} Initialize the Lagrangian variables $\lambda_{k}^{(0)}$, $\mu_{k}^{(0)}$, $\omega^{(0)}$, $\chi^{(0)}$, $\forall k$;
\STATE \vspace{0.1cm}{\bf Else}
\STATE \vspace{0.1cm}\hspace{0.35cm} {\bf While} $C_{2}^{(i+1)}-C_{2}^{(i)}\geq \Delta$ or $i\leq i_{\rm{max}}$
\STATE \vspace{0.1cm}\hspace{0.35cm} {\bf do}
\STATE \vspace{0.1cm}\hspace{0.35cm} Calculate $B_{k}^{(i+1)}$ and $P_{2}(h_{k})^{(i+1)}$ using Eq.~\eqref{eq:Bk_i+1};
\STATE \vspace{0.1cm}\hspace{0.35cm} Calculate $C_{2}^{(i+1)}$ with $B_{k}^{(i+1)}$ and $P_{2}(h_{k})^{(i+1)}$ using Eq.~\eqref{eq:C2};
\STATE \vspace{0.1cm}\hspace{0.35cm} Update $\lambda_{k}^{(i+1)}$, $\mu_{k}^{(i+1)}$, $\omega^{(i+1)}$, $\chi^{(i+1)}$ using Eq.~\eqref{eq:Bk_i+1};
\STATE \vspace{0.1cm}\hspace{0.35cm} Set $i=i+1$;
\STATE \vspace{0.1cm}\hspace{0.35cm} {\bf End while}
\STATE \vspace{0.1cm}\hspace{0.35cm} Set $C_{2\rm{max}}\triangleq C_{2}^{(i+1)}$;
\STATE \vspace{0.1cm}\hspace{0.35cm} Set $[B_{k}^{*}, P_{2}(h_{k})^{*}]\triangleq [B_{k}^{(i+1)}, P_{2}(h_{k})^{(i+1)}]$;
\STATE {\bf End if}
\end{algorithmic}
\end{algorithm}

\subsection{Energy Efficiency Analysis}
The power consumption of circuit blocks is given by~\cite{1321221}\\
\begin{eqnarray}
\begin{aligned}
&P_C=M_t(P_{DAC}+P_{MIX}+P_{FILT})+2P_{SYN}\\
&\hspace{0.5cm}+M_r(P_{LNA}+P_{MIX}+P_{IFA}+P_{FILR}+P_{ADC})
\end{aligned}\label{27}
\end{eqnarray}
where $M_t$ and $M_r$ are the numbers of transmit and receive antennas, respectively. $P_{DAC}$, $P_{MIX}$, $P_{LNA}$, $P_{IFA}$, $P_{FILT}$, $P_{FILR}$, $P_{ADC}$, and $P_{SYN}$ are power consumptions for DAC, mixer, low-noise amplifier (LNA), intermediate frequency amplifier (IFA), active filters at the transmit side, active filters at the receive side, ADC, and frequency synthesizer, respectively. To estimate the values of $P_{DAC}$ and $P_{ADC}$, we use the model introduced in~\cite{1532220}.

The power consumption of relay, denoted by $P_{RELAY}$, is given by\\
\begin{equation}
\label{28}
P_{RELAY}=\frac{P_C+P_{FPGAR}+P_{PA}}{\eta},
\end{equation}
where $P_C$, $P_{FPGAR}$, and $P_{PA}$ are the power consumption values for all circuit blocks, the FPGA on relay, and the power amplifier, respectively. $\eta$ is the power conversion efficiency. On the other hand, the power consumption of IRS, denoted by $P_{IRS}$, can be given as follows:\\
\begin{equation}
\label{29}
P_{IRS}=\frac{P_{FPGAI}+N \times P_{PIN}}{\eta},
\end{equation}
where $P_{FPGAI}$ and $P_{PIN}$ are the power consumptions values for the FPGA on IRS and the PIN diode, respectively. \\
%
%


For the relaying systems, $\gamma_{k,1}$ denotes the instantaneous SNR from the on-site command center to the $k$-th 5G-UAV relay, $\gamma_{k,2}$ denotes the instantaneous SNR from the $k$-th relay to the trapped user.
Then, $\gamma_{k,1}$ and $\gamma_{k,2}$ can be expressed as follows:
\begin{eqnarray}
\left\{\begin{array}{ll}
\vspace{0.2cm}\gamma_{k,1} =\dfrac{P_{S}}{\delta^{2}}(L_{k}^{SR})^{-\alpha};\\
\vspace{0.2cm}\gamma_{k,2} =\dfrac{P_{PA}}{\delta^{2}}(L_{k}^{RD})^{-\alpha}g_{k},
\end{array}\right.
\end{eqnarray}
where $P_{S}$ is the transmit power from the on-site command center. The Fisher-Snedecor $\mathcal{F}$ RVs $g_{k}$ is modelled as independently distributed with the PDF given by Eq.~\eqref{eq:fgkn-pdf}. Since there are mutual interferences among relays when the bandwidth is shared, we sperate the available bandwidth as $K$ orthogonal sub-channels. The ergodic capacity for relaying systems, denoted by $C_{RELAY}$, can be derived as follows:
\begin{align}
\label{eq:RELAY_capacity}
&C_{RELAY}\!\!=\!\!\underbrace{\iint\ldots\int_{0}^{\infty}}_{K-fold}\sum_{k=1}^{K} \dfrac{B}{2K}\log_{2}\left(1\!+\!\dfrac{4\gamma_{k,1}\gamma_{k,2}}{1\!+\!2\gamma_{k,1}\!+\!2\gamma_{k,2}}\right)\nonumber\\
&\hspace{4.5cm}\times\prod\limits_{k=1}^{K}f_{k,n}(g_{k})\mathrm{d}g_{1}\ldots\mathrm{d}g_{K}.
\end{align}

Then, we have the energy efficiencies of relay and IRS, denoted by $\varrho_{1}$ and $\varrho_{2}$, respectively, as follows:
\begin{eqnarray}
\left\{\begin{array}{ll}
\vspace{0.2cm}\!\!\varrho_{1}\!\!=\!\dfrac{C_{RELAY}}{K(P_{RELAY}+P_{S}/\eta)},&\hspace{-0.2cm}\text{for 5G-UAV with relay};\\
\vspace{0.2cm}\!\!\varrho_{2}\!\!=\!\dfrac{C_{IRS}}{K (P_{IRS}+P_{S}/\eta)},&\hspace{-0.2cm}\text{for 5G-UAV with IRS},\\
\end{array}\right.\label{EE}
\end{eqnarray}
where $C_{RELAY}$ and $C_{IRS}$ are the ergodic capacities of 5G-UAV with relay and 5G-UAV with IRS under the heterogeneous Fisher-Snedecor $\mathcal{F}$ composite fading channels for 5G-UAV based EWC networks. Since the bandwidth is divided into $K$ sub-channels in the relaying based 5G-UAV networks, we consider the ergodic capacity of IRS based 5G-UAV networks under the same case. Then, $C_{IRS}$ can be rewritten as Eq.~\eqref{eq:C_irs}.

\begin{figure*}
\begin{equation}
\label{eq:C_irs}
\begin{split}
C_{IRS} = \underbrace{\iint\ldots\int_{0}^{\infty}}_{K-fold}\displaystyle\sum_{k=1}^{K} \dfrac{B}{K}\log_{2}
\left\{1+\text{min}\left\{\left[\dfrac{\bar{P}_{h}+\frac{\epsilon \Lambda N_{0}B}{K^{2}}\prod\limits_{k=1}^{K}L_{k}^{\alpha}}{\prod\limits_{k=1}^{K}L_{k}^{\alpha}}-\dfrac{N_{0}B}{Kh}\right]^{+},~P_{h}\right\}
\dfrac{Kh}{N_{0}B}\right\}\prod\limits_{k=1}^{K}f_{k}(h_{k})\mathrm{d}h_{1}\ldots\mathrm{d}h_{K}.
\end{split}
\end{equation}
\hrulefill
\end{figure*}

Although the final analytical result in Eq.~\eqref{eq:C_irs} seems to be complicated, numerical results can be obtained by the folded accumulation and summation of the integral variables. The item $\left[\frac{\bar{P}_{h}+\frac{\epsilon \Lambda N_{0}B}{K^{2}}\prod\limits_{k=1}^{K}L_{k}^{\alpha}}{\prod\limits_{k=1}^{K}L_{k}^{\alpha}}-\frac{N_{0}B}{Kh}\right]^{+}$ is the allocated power corresponding to the instantaneous power gain $h_{k}$, which is obtained by the optimal solution of ${\bf \emph{P}1}$ specified in Eq.~\eqref{eq:power_C1}. For each accumulation, the allocated power is compared with the peak value of available power and the smaller one is adopted.

For energy efficiency aspect, the superiority of IRS-assisted 5G-UAVs over the traditional relays is mainly due to the low energy consumption of PIN diodes. The power consumption of traditional relays comes from all circuit blocks, the FPGA on the relay, and the power amplifier, which are specified in Eq.~\eqref{28}. The power consumption of IRS-assisted 5G-UAVs comes from the FPGA on IRS and the PIN diodes, which are specified in Eq.~\eqref{29}. The total power consumptions with IRS cases are very low since there are no RF chains which are replaced by the PIN diodes with low power consumptions. For the resource limited emergency wireless communication, it is very important to use the light-weight and low power PIN diodes in such case sensitive communication scenarios.
\section{Performance Evaluations}\label{sec:simu}

\begin{figure}[h]
\centering
\includegraphics[scale=0.5]{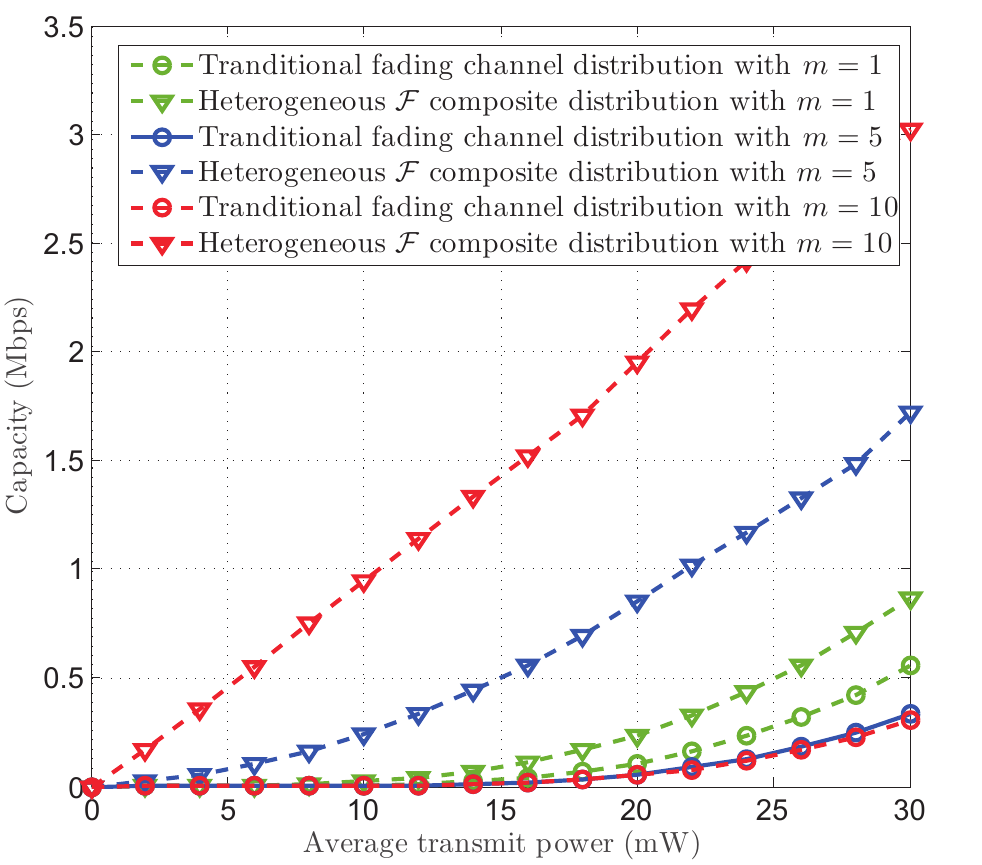}
\caption{The comparison of channel capacity between the new heterogeneous $\mathcal{F}$ composite fading channel adapted power allocation scheme and the traditional channel model adapted power allocation scheme with fading parameters.} \label{channel_capacity}
\end{figure}

\begin{figure}[h]
\centering
\includegraphics[scale=0.5]{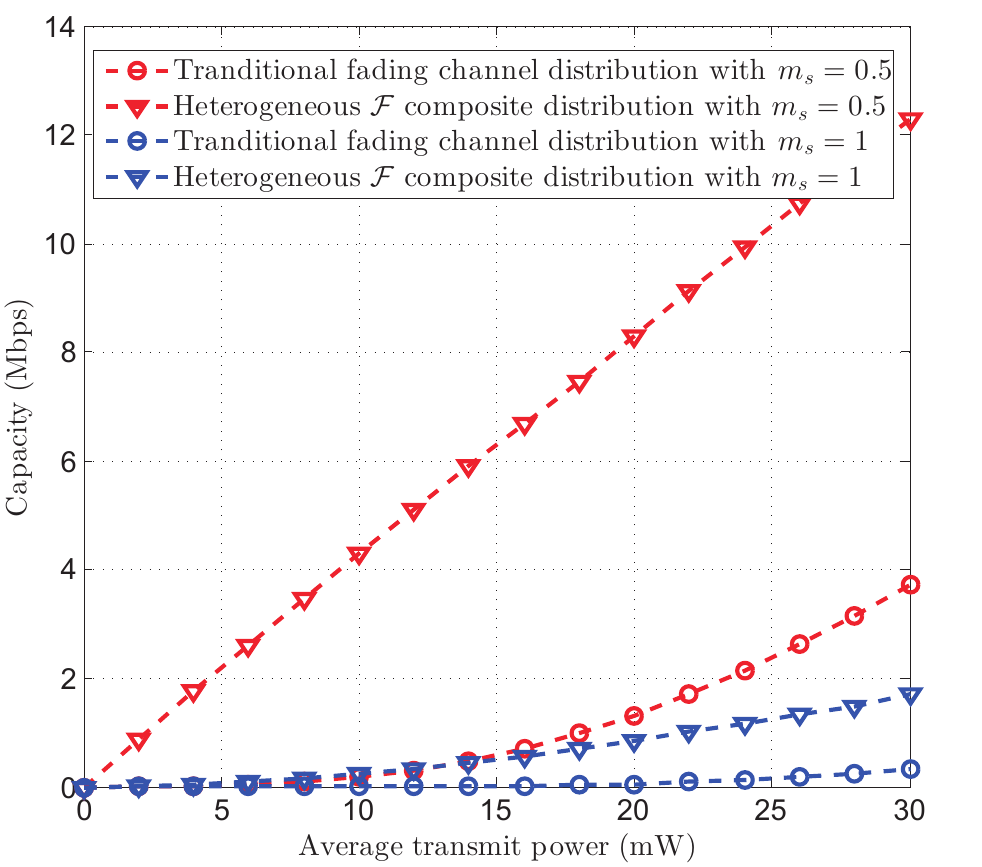}
\caption{The comparison of channel capacity between the new heterogeneous $\mathcal{F}$ composite fading channel adapted power allocation scheme and the traditional channel model adapted power allocation scheme with shadowing parameters.} \label{channel_capacity_ms}
\end{figure}

\begin{figure}[h]
\centering
\includegraphics[scale=0.5]{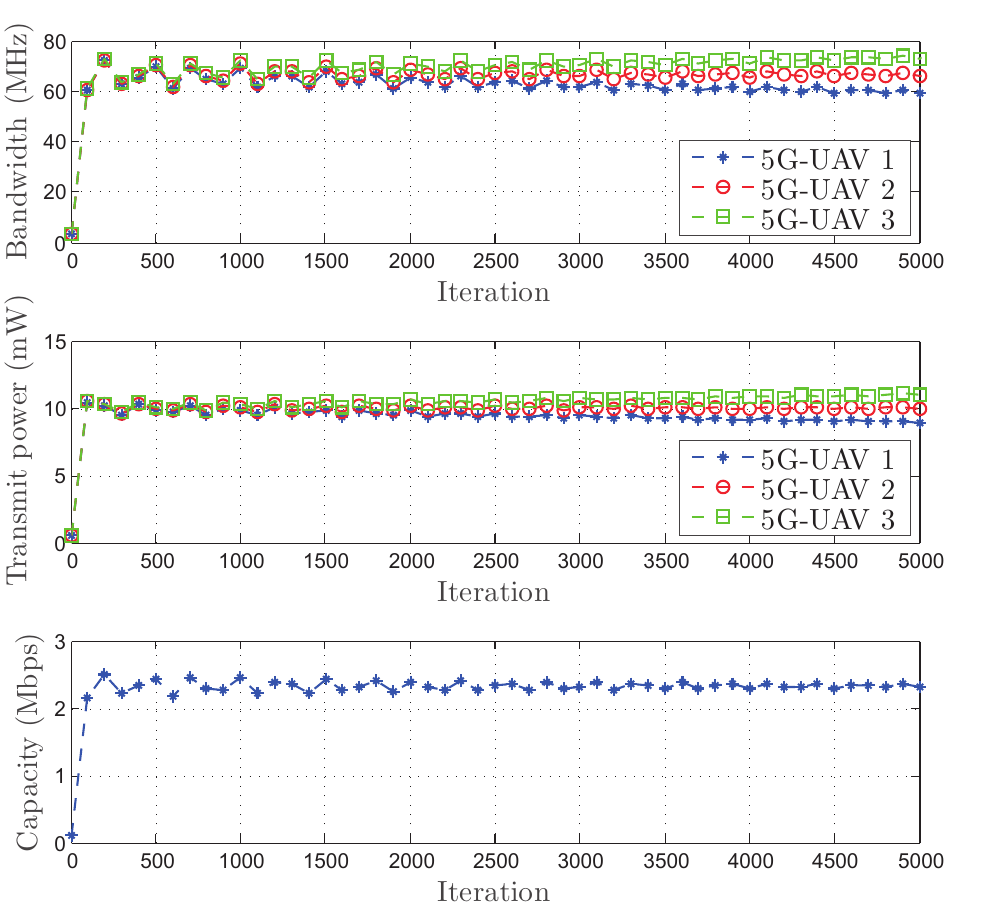}
\caption{Joint bandwidth-power allocation scheme with very close values for power gains of $K$ sub-channels ($h_{1}=5.1, h_{2}=5.2$, and $h_{3}=5.3$).} \label{Joint-opt--8000}
\end{figure}

\begin{figure}[h]
\centering
\includegraphics[scale=0.5]{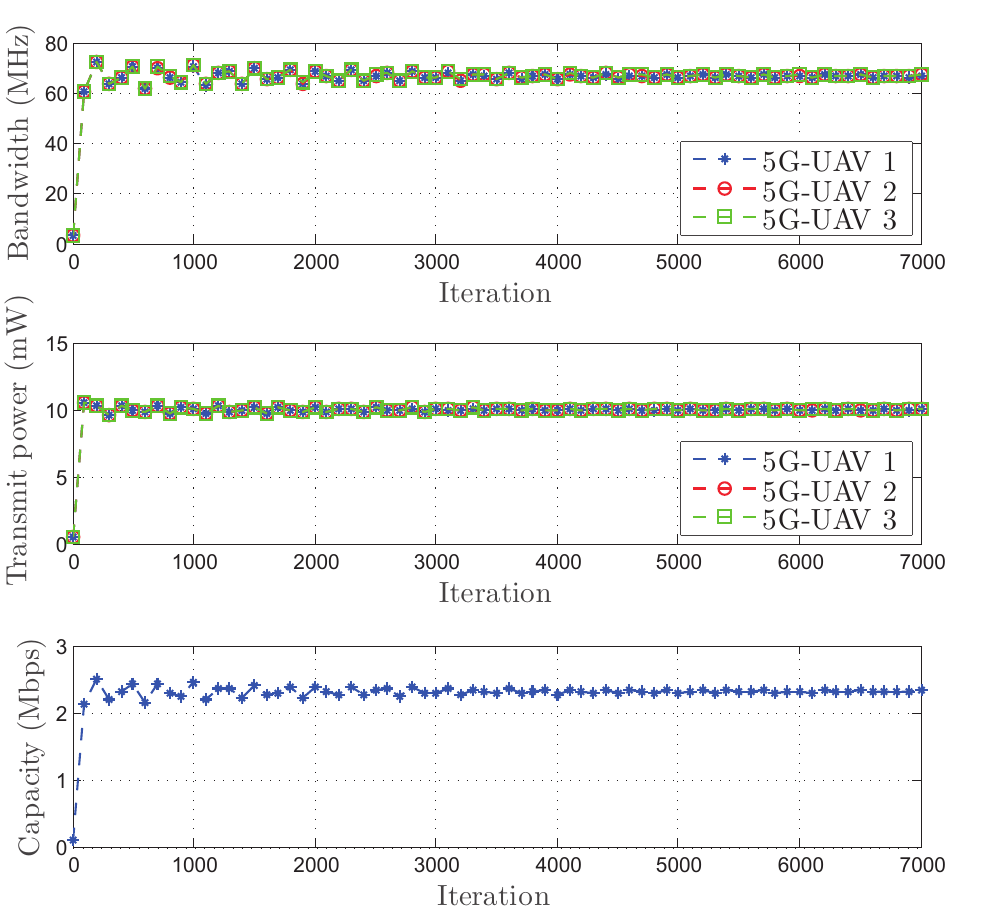}
\caption{Joint bandwidth-power allocation scheme with the same value for power gains of $K$ sub-channels ($h_{1}=h_{2}=h_{3}=5$).} \label{Joint-opt--same-h-8000}
\end{figure}

\begin{figure}[h]
\centering
\includegraphics[scale=0.5]{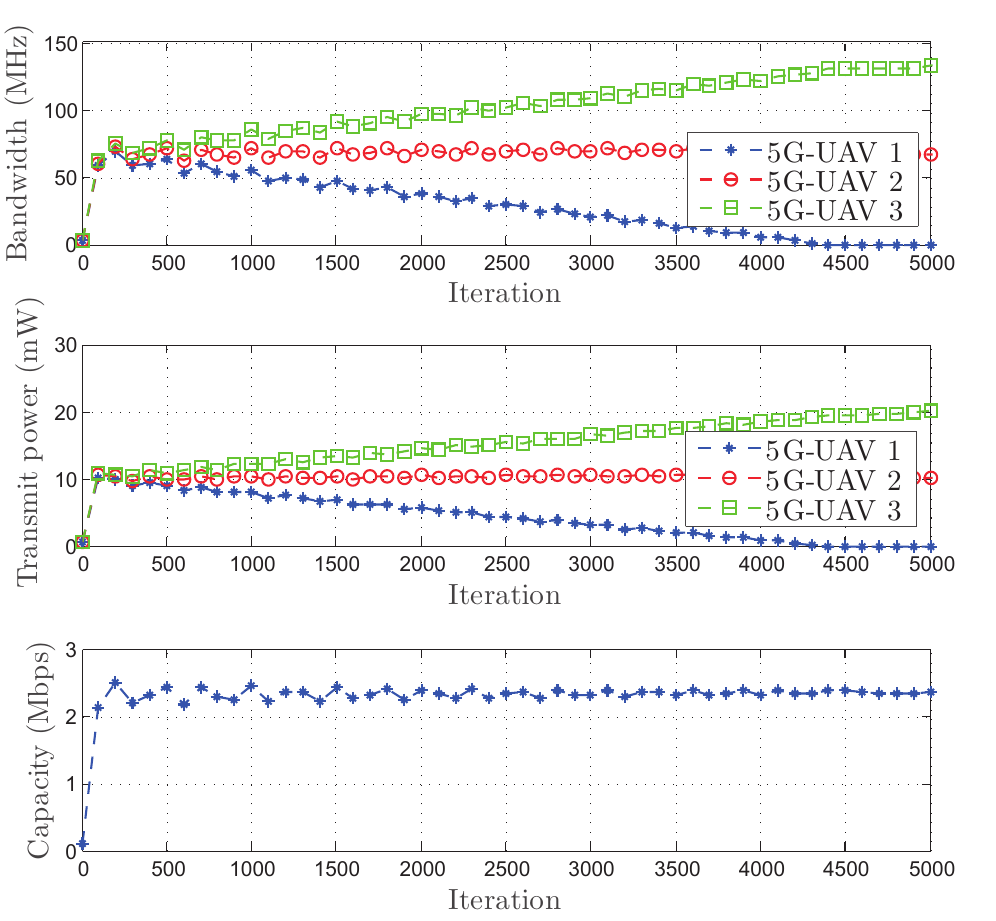}
\caption{Joint bandwidth-power allocation scheme with the diverse values for power gains of $K$ sub-channels ($h_{1}=4, h_{2}=5$, and $h_{3}=6$).} \label{Joint-opt--0--8000}
\end{figure}

\begin{figure}[h]
\centering
\includegraphics[scale=0.5]{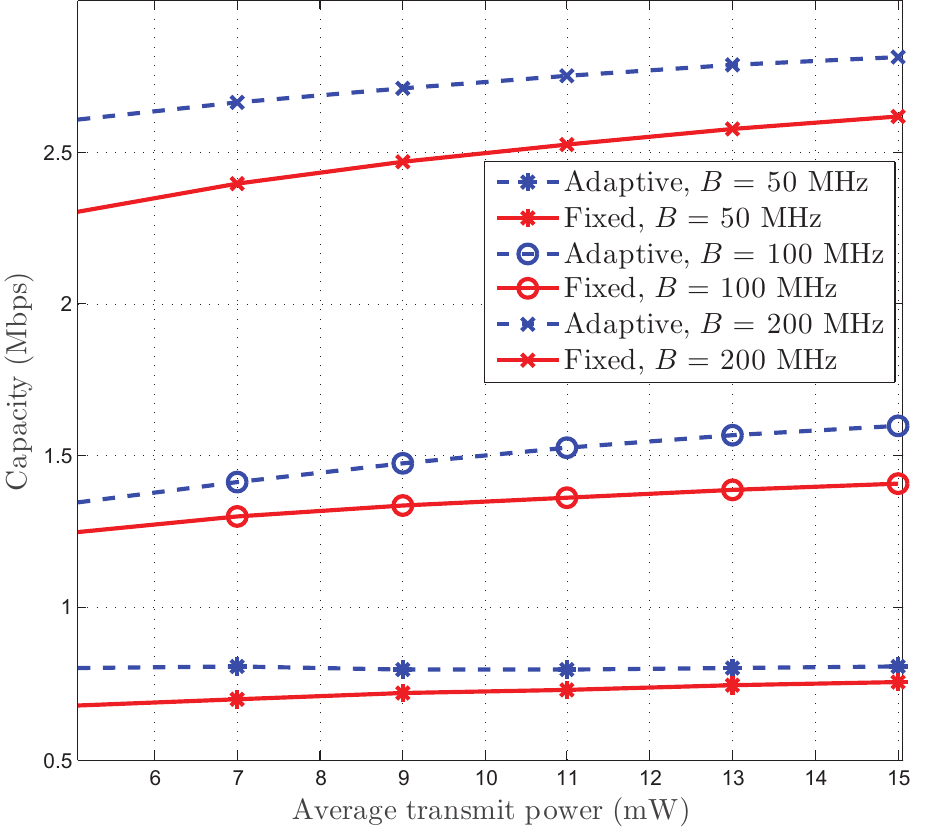}
\caption{The capacities with the joint bandwidth-power allocation scheme and fixed-bandwidth based power allocation scheme.} \label{capacity-A-F}
\end{figure}

\begin{figure}[h]
\centering
\includegraphics[scale=0.5]{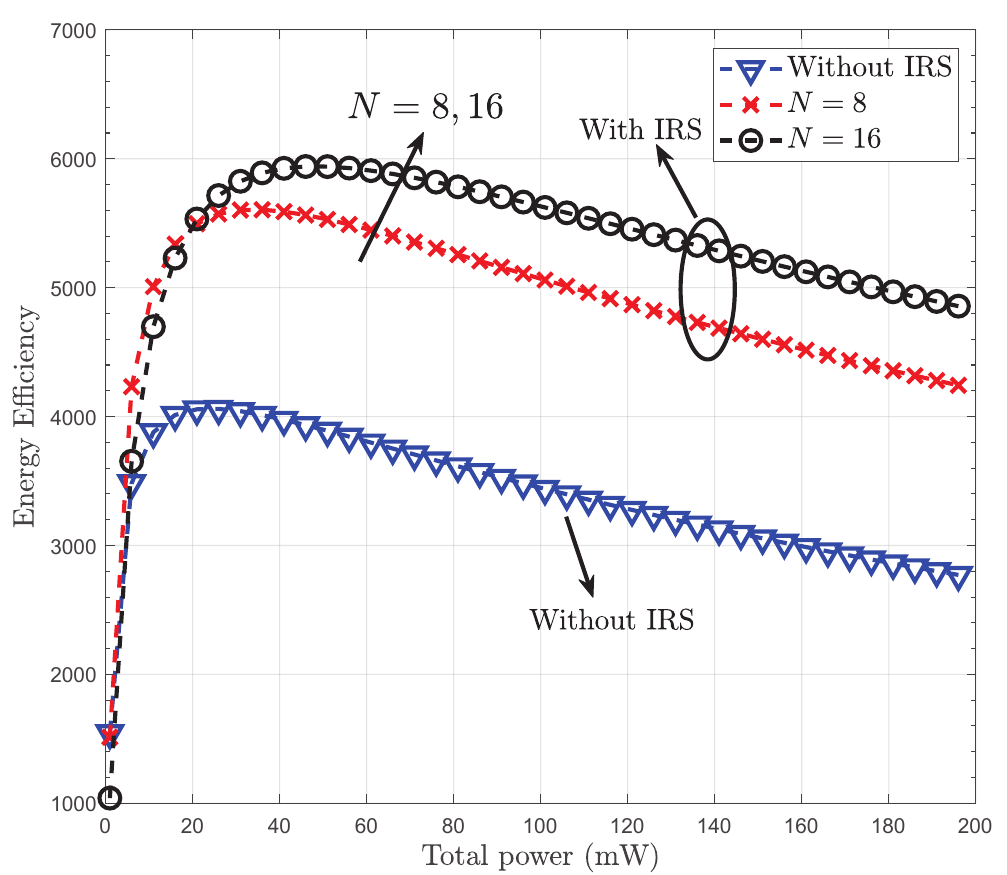}
\caption{The energy efficiency versus the total power for 5G-UAVs with and without IRSs.} \label{IRS-PIN}
\end{figure}

\begin{figure}[h]
\centering
\includegraphics[scale=0.5]{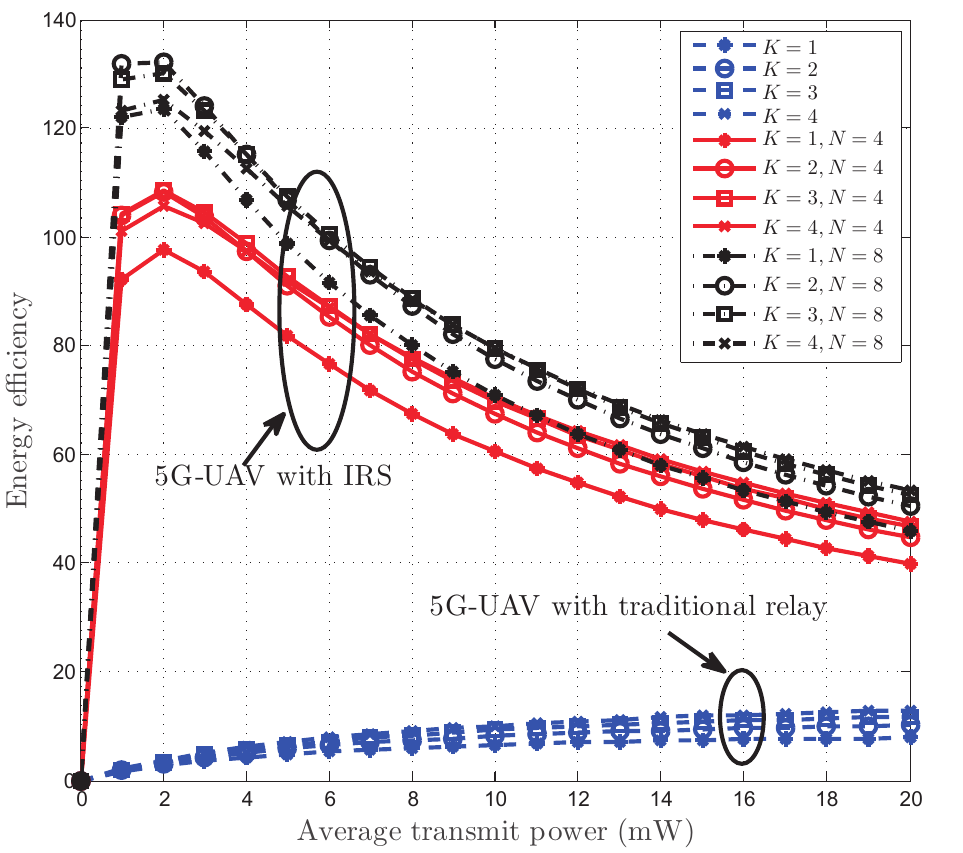}
\caption{The comparison of energy efficiency between 5G-UAVs with IRSs and the traditional relays.} \label{EE-performance}
\end{figure}

\begin{figure}
\centering
\includegraphics[scale=0.52]{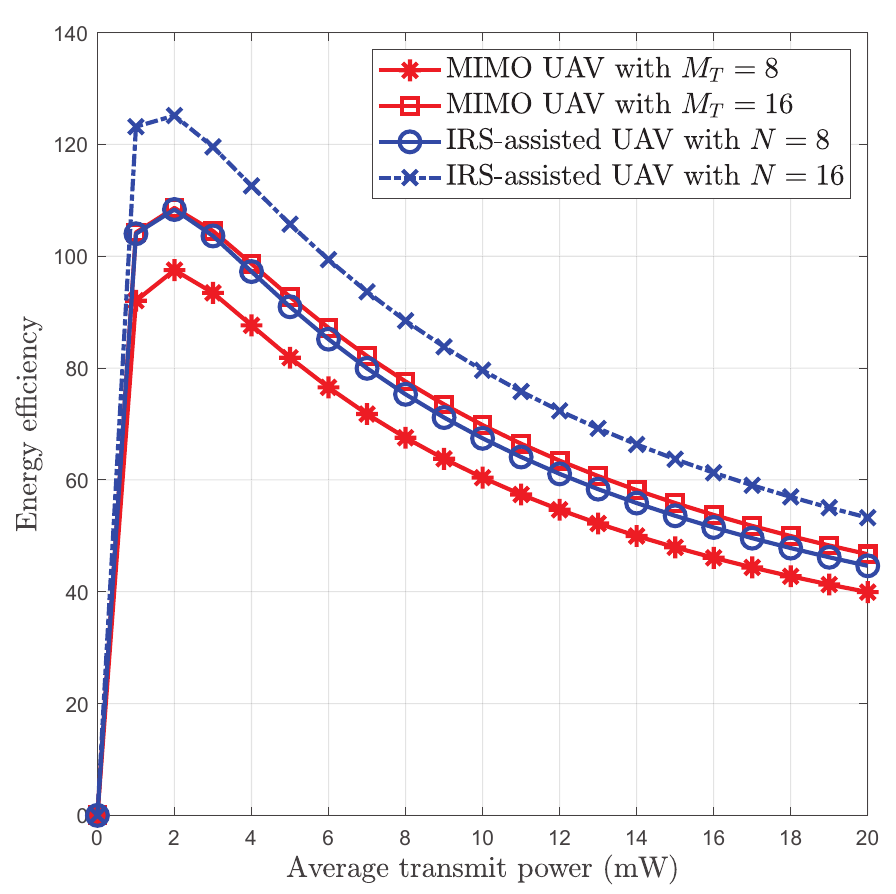}
\caption{The comparison of energy efficiency between IRS-assisted UAVs and MIMO UAVs.} \label{IRS-MIMO}
\end{figure}

In this section, we evaluate the performances of capacity and energy efficiency enhancement with numerical results. First, we compare the capacity corresponding to heterogeneous $\mathcal{F}$ composite fading channel adapted optimal power allocation scheme and the capacity of traditional fading channel distribution (like Rayleigh distribution) adapted scheme with average transmit power in Figs.~\ref{channel_capacity} and~\ref{channel_capacity_ms}. Next, we evaluate the joint bandwidth-power allocation algorithm, and analyze the convergence under different power gains, as shown in Figs.~\ref{Joint-opt--8000},~\ref{Joint-opt--same-h-8000}, and~\ref{Joint-opt--0--8000}. Fig.~\ref{capacity-A-F} shows the comparisons between joint bandwidth-power allocation algorithm (i.e., adaptive bandwidth allocation) and the fixed bandwidth allocation scheme. Furthermore, since the power resource is focused constraints under EWC networks, we show the advantage of energy efficiency for IRS-assisted 5G-UAVs in Figs.~\ref{IRS-PIN}, ~\ref{EE-performance}, and~\ref{IRS-MIMO}. We employ the heterogeneous $\mathcal{F}$ composite fading channel model with the parameter settings shown in TABLE \uppercase\expandafter{\romannumeral 1}.

\begin{table*}[!t]
\renewcommand{\arraystretch}{1.3}
\caption{The Main Parameters For Simulation}
\label{table_example}
\centering
\begin{tabular}{|c|c|c|}
\hline
\bfseries Symbol & \hspace{0.8cm}\bfseries Value \hspace{0.8cm} & \hspace{1.5cm}\bfseries Definition \hspace{1.5cm} \\
\hline
$K$ & 1, 2, 3, 4   & The number of 5G-UAVs/sub-channels\\
\hline
$B$ & 200 MHz   & The total available bandwidth\\
\hline
$P_{h}$ &$0\sim1\rm{W}$   & The peak transmit power\\
\hline
$\beta$ & 0.01 & The step factor for the algorithm \\
\hline
$L_{k}$ &$100\rm{m}\sim500\rm{m}$ &The average distance from on-site command center to the trapped user at the $k$-th sub-channel\\
\hline
$\alpha$ &$0\sim 1$ &The path-loss exponent\\
\hline
$\bar{h}_{k}/ \bar{g}_{k}$ &5/1 &The average power gain of the $k$-th 5G-UAV/relay sub-channel  \\
\hline
$m$ &$1\sim20$ &The fading parameter\\
\hline
$m_{s}$ &$1\sim20$ &The shadowing parameter\\
\hline
$i_{max}$ &8000 & The number of iterations for the algorithm\\
\hline
$\triangle$ &0.01 &The stopping threshold for the algorithm\\
\hline
$P_{S}$ &$0\sim1\rm{W}$   & The transmit power from the on-site command center\\
\hline
$P_{MIX}$ & 30.3mW &The power consumption for the mixer\\
\hline
$P_{FILT}$&2.5W &The power consumption for the active filters at the transmitter side\\
\hline
$P_{FILR}$&2.5W &The power consumption for the active filters at the receiver side\\
\hline
$P_{SYN}$&50mW & The power consumption for the frequency syntherizer\\
\hline
$P_{LNA}$&20mW &The power consumption for LNA\\
\hline
$P_{IFA}$&3mW &The power consumption for IFA\\

\hline
$P_C$&3W & The power consumption for all circuit blocks\\
\hline
$P_{FPGAR}$&1W &The power consumption for FPGA on relay\\
\hline
$P_{PA}$&5W &The power consumption for power amplifier\\
\hline
$P_{FPGAI}$&0.5W &The power consumption for FPGA on IRS\\
\hline
$P_{PIN}$&8.5mW &The power consumption for PIN diode\\
\hline
$\eta$&80\% &The power conversion efficiency\\
\hline
$N$&$4, 8, 16$ &The number of IRS reflective elements\\
\hline
\end{tabular}
\end{table*}


Figure~\ref{channel_capacity} shows the comparison of channel capacity between the new heterogeneous $\mathcal{F}$ composite fading channel adapted power allocation scheme given in Section~\ref{optimization}-B and the traditional channel model (like Rayleigh distribution) adapted power allocation scheme with fading parameters, where the shadowing parameter is set to $m_{s}=1$. The fading parameters are set to $m=1$, $m=5$, and $m=10$, respectively. It is clear that the heterogeneous $\mathcal{F}$ composite fading channel adapted power allocation scheme outperforms the traditional fading channel adapted power allocation scheme. The ergodic capacity of the heterogeneous $\mathcal{F}$ composite fading channel adapted optimal power allocation scheme increases as fading parameter $m$ increases. However, the ergodic capacity of the traditional channel model (like Rayleigh distribution) adapted power allocation scheme slightly increases only at $m=1$. The curves of ergodic capacities at $m=5$ and $m=10$ are almost coincide with a low value. This is because the traditional power allocation scheme does not consider the variation of fading and shadowing.

Figure~\ref{channel_capacity_ms} shows the comparison of channel capacity between the new heterogeneous $\mathcal{F}$ composite fading channel adapted power allocation scheme given in Section~\ref{optimization}-B and the traditional channel model (like Rayleigh distribution) adapted power allocation scheme with shadowing parameters, where the fading parameter is set to $m=5$. The shadowing parameters are set to $m_{s}=0.5$ and $m_{s}=1$, respectively. We can find that the ergodic capacity increases as the shadowing parameter decreases since the channel at a small $m_{s}$ is allocated much power shown in Fig.~\ref{power_allocation_C1}. Clearly, the heterogeneous $\mathcal{F}$ composite fading channel adapted optimal power allocation scheme performs much better than the traditional channel model adapted power allocation scheme to obtain the maximum ergodic capacity for EWC networks.


Then, we evaluate the performance of our proposed joint bandwidth-power allocation scheme for multiple 5G-UAV based EWC networks given in Section~\ref{optimization}-C with Figs.~\ref{Joint-opt--8000},~\ref{Joint-opt--same-h-8000}, and~\ref{Joint-opt--0--8000}, where the number of 5G-UAVs $K$ is set to $3$. The available bandwidth is set to $200$MHz, and the available power is set to $30$mW. The average power gain of the heterogeneous $\mathcal{F}$ composite fading channel is set to $5$. We depict the bandwidth, power, and capacity convergence with different power gains, which illustrates the fast convergence with our proposed joint bandwidth-power allocation scheme. Fig.~\ref{Joint-opt--8000} shows the resource allocation strategy with very close values for power gains of $K$ sub-channels, where $h_{1}=5.1, h_{2}=5.2$, and $h_{3}=5.3$. As shown in Fig.~\ref{Joint-opt--8000}, the allocated bandwidth and power resources for different 5G-UAVs all converge to fixed values around the iteration number of $4500$~(For example, the allocated bandwidth for 5G-UAV $1$ converges to the value around $60$MHz and the transmit power for 5G-UAV $1$ converges to the value around $9$mW).

Figure~\ref{Joint-opt--same-h-8000} shows the joint bandwidth-power allocation scheme with the same quality of $K$ sub-channels, where $h_{1}=h_{2}=h_{3}=5$. As illustrated in Fig.~\ref{Joint-opt--same-h-8000}, the allocated bandwidth for each 5G-UAV converges to the value around $65$MHz and the transmit power for each 5G-UAV converges to the value around $10$mW.

Figure~\ref{Joint-opt--0--8000} shows the joint bandwidth-power allocation scheme with different qualities of $K$ sub-channels, where $h_{1}=4, h_{2}=5$, and $h_{3}=6$. To obtain the maximum capacity, the best sub-channel with $h_{3}=6$ is allocated the most bandwidth-power resources, while the worst sub-channel with $h_{1}=4$ is allocated the least. Around the iteration number of $4500$, the allocated bandwidth and transmit power resources for 5G-UAV $1$ both converge to zero. In addition, the maximum capacities in Figs.~\ref{Joint-opt--8000}, ~\ref{Joint-opt--same-h-8000}, and~\ref{Joint-opt--0--8000} all converge to fixed values with the optimal resource allocation scheme.

Figure~\ref{capacity-A-F} shows the capacities with the joint bandwidth-power allocation scheme and fixed-bandwidth based power allocation scheme. As illustrated in Fig.~\ref{capacity-A-F}, the adaptive resource allocation algorithm, i.e., the joint bandwidth-power allocation scheme, can achieve higher capacity than the fixed-bandwidth based power allocation scheme. This is because the allocated bandwidth with joint bandwidth-power allocation scheme is adaptive to the instantaneous channel state, which can be validated from Figs.~\ref{Joint-opt--8000},~\ref{Joint-opt--same-h-8000}, and~\ref{Joint-opt--0--8000}.

To clarify the impact of active PIN diodes on the energy efficiency of IRS-assisted 5G-UAVs for emergency wireless communication scenarios, we depict the energy efficiency versus the total power for 5G-UAVs with and without IRSs, as shown in Fig.~\ref{IRS-PIN}, where $N$ is the number of reflectors corresponding to active PIN diodes. We set $N$ as 8 and 16, respectively. Numerical results show that the energy efficiency of IRS-assisted 5G-UAVs is higher than that of 5G-UAVs without IRSs under given power constraint.


Figure~\ref{EE-performance} plots energy efficiencies under the multiple 5G-UAVs with IRSs cooperation network and multiple relays cooperation network, respectively, where the numbers of 5G-UAVs are set to $K=1, K=2, K=3$, and $K=4$, respectively. For 5G-UAV with IRSs, the numbers of reflective elements are set to $N=4$ and $N=8$, respectively. As shown in Fig.~\ref{EE-performance}, the energy efficiency firstly increases and gradually decreases as the average transmit power increases under the IRS based 5G-UAVs cooperation network while the energy efficiency of relay based 5G-UAVs cooperation network gradually increases to be fixed as the average transmit power increases. Also, the energy efficiency changes with different numbers of 5G-UAV $K$. The energy efficiency increases as the number of 5G-UAVs increases. In addition, Since the power consumption of PIN is extremely small while the power gain significantly increases, the energy efficiencies of 5G-UAVs with IRSs increase as the number of reflective elements increases. In general, the energy efficiency of IRS based 5G-UAV is higher than that of relay based 5G-UAV.

To further reveal the system insights of our proposed IRS-assisted 5G-UAV system for emergency wireless communication scenarios, Fig.~\ref{IRS-MIMO} compares the energy efficiency of IRS-assisted UAVs and the energy efficiency of MIMO UAVs, where $M_{T}$ is the number of antennas corresponding to MIMO UAV and $N$ is the number of reflectors corresponding to IRS-assisted UAV. Numerical results show that the energy efficiency of IRS-assisted UAVs is significantly higher than that of MIMO UAVs with the same numbers of antennas. This is mainly because each reflector in IRS-assisted UAV is smartly controlled by a corresponding PIN diode, the power consumption of which is much smaller than that of power amplifier in MIMO UAV, specified in Eqs.~\eqref{28} and~\eqref{29}.

\section{Conclusions}\label{sec:conc}
In this paper, we investigated the 5G-UAV based EWC. In particular, we propose the heterogeneous $\mathcal{F}$ composite fading channel model in EWC networks, which provides the accurate model and characterization with the path-loss, the number of reflectors, fading, and shadowing, for the complex communication scenarios. Based on the channel model, we develop the optimal power allocation scheme and joint bandwidth-power allocation scheme to enhance capacity and energy efficiency performances for EWC networks. Numerical results show that the capacity of the heterogeneous $\mathcal{F}$ composite fading channel adapted power allocation scheme is higher than that of other traditional fading channel adapted power allocation schemes. Also, the well convergence of the joint bandwidth-power allocation scheme is revealed, which outperforms other fixed-bandwidth based power allocation scheme. In addition, we compare the energy efficiencies between IRS and relay based 5G-UAV for EWC networks, showing that the 5G-UAV with IRS is superior to the traditional relays in the aspect of energy efficiency, thus being conducive to restore the communications in post-disaster areas.

\section*{Appendix A \\Proof of Lemma 1}\label{appendix_a}
\emph{Proof}: According to Eq.~\eqref{eq:fgkn-pdf}, the PDF of $g_{k}$, denoted by $f_{g}(g_{k})$, can be obtained as follows~\cite{8359199}:
\begin{eqnarray}
\begin{aligned}
f_{g}(g_{k})&= \frac{g_{k}^{Nm_{k}-1}}{B (Nm_{k}, Nm_{s_{k}})}\left(\frac{m_{k}}{Nm_{s_{k}} \bar{g}_{k}}\right)^{Nm_{k}} \\
& \times{}_{2} F_{1}\left(N(m_{k}+m_{s_{k}}), Nm_{k}; Nm_{k}; \frac{-m_{k}g_{k}}{Nm_{s_{k}} \bar{g}_{k}} \right).
\end{aligned}\label{eq:gk_pdf}
\end{eqnarray}
The function ${ }_{2} F_{1}(\alpha, \beta; \gamma; -z)$ is the hypergeometric function~\cite[Eq.(9.111)]{2007859}, which is given as follows:
\begin{eqnarray}
{ }_{2} F_{1}(\alpha, \beta; \gamma; -z)=\frac{\Gamma(\gamma) z}{\Gamma(\alpha) \Gamma(\beta)} G_{2,2}^{1,2}\left(z \bigg| \begin{array}{c}
-\alpha,-\beta \\
-1,-\gamma
\end{array}\right),
\label{F-function}
\end{eqnarray}
where $G_{p,q}^{m,n}\left(z \bigg| \begin{array}{c}
a_{1}, \cdots, a_{p}\\
b_{1}, \cdots, b_{q}
\end{array}\right)$ is the Meijer's G-function~\cite[Eq.(9.301)]{2007859}.
Using Eq.~\eqref{F-function}, the PDF specified by Eq.~\eqref{eq:gk_pdf} can be rewritten as Eq.~\eqref{eq:gk_pdf2}.
\begin{figure*}
\begin{equation}
\label{eq:gk_pdf2}
\begin{split}
f_{g}(g_{k})&=\frac{g_{k}^{Nm_{k}}}{\Gamma(Nm_{k}) \Gamma\left(N m_{s_{k}}\right)}\left(\frac{m_{k}}{Nm_{s_{k}}\bar{g}_{k}}\right)^{Nm_{k}+1} G_{2,2}^{1,2}\left(\frac{m_{k}g_{k}}{Nm_{s_{k}}\bar{g}_{k}} \bigg|\!\! \begin{array}{c}
-N(m_{k}\!+\!m_{s_{k}}),-N m_{k} \\
-1,-N m_{k}
\end{array}\right).
\end{split}
\end{equation}
\hrulefill
\begin{equation}
\label{eq:hk_PDF}
\begin{split}
f_{k}(h_{k})=\frac{\partial F_{k}(h)}{\partial h}\bigg|_{h_{k}}= L_{k}^{\alpha}f_{g}(h_{k}L_{k}^{\alpha})=\frac{L_{k}^{\alpha}\Lambda_{k}}{\Gamma(Nm_{k}) \Gamma\left(N m_{s_{k}}\right)} G_{2,2}^{1,2}\left(\Lambda_{k}h_{k} \bigg|\begin{array}{c}
-Nm_{s_{k}},0 \\
N m_{k}-1,0
\end{array}\right),
\end{split}
\end{equation}
\hrulefill
\end{figure*}

The cumulative distribution function (CDF) of $g_{k}$ can be expressed as $F_{g}(g)=\int_{0}^{g}f_{g}(g_{k})\mathrm{d} g_{k}$. Due to $g_{k}=L_{k}^{\alpha}h_{k}$, the CDF of $h_{k}$ can be expressed as $F_{k}(h)=\int_{0}^{L_{k}^{\alpha}h}f_{g}(g_{k})\mathrm{d} g_{k}$. Then, we can obtain the PDF of $h_{k}$ specified by Eq.~\eqref{eq:hk_PDF} and Lemma 1 follows. $\hfill\blacksquare$

\section*{Appendix B \\Proof of Lemma 2}\label{appendix_B}
\emph{Proof}: To obtain the PDF of heterogeneous $\mathcal{F}$ composite fading channel, the moment generating function (MGF) of $h_{k}$, denoted by $\mathcal{M}_{h_{k}}(t)$, with the help of~\cite[Eq.(7.811.5)]{2007859} and~\cite[Eq.(9.31.2)]{2007859}, can be expressed as Eq.~\eqref{eq:MGF-hk}, where the MGF is given by $\mathcal{M}(t) \triangleq \int_{0}^{\infty} \exp (-x t) f_{X}(x) d x$.

Then, the MGF of $h$ can be written as
\begin{eqnarray}
\mathcal{M}_{h}(t)=\prod_{k=1}^{K} \mathcal{M}_{h_{k}}(t).
\label{eq:MGF-h}
\end{eqnarray}
\\
Based on Eq.~\eqref{eq:MGF-hk} and the definition of the Meijer's G-function~\cite[Eq.(9.301)]{2007859}, Eq.~\eqref{eq:MGF-h} can be rewritten as Eq.~\eqref{eq:MGF-t}.
\begin{figure*}
\begin{equation}
\label{eq:MGF-hk}
\begin{split}
\mathcal{M}_{h_{k}}(t) =\frac{L_{k}^{\alpha}}{\Gamma\left(Nm_{k}\right) \Gamma\left(Nm_{s_{k}}\right)} (\frac{\Lambda_{k}}{t})^{Nm_{k}}
\times \mathrm{G}_{3,2}^{1,3}\left(\frac{\Lambda_{k}}{t}\bigg| \begin{array}{c} 1\!\!-\!\!Nm_{k},1\!\!-\!\!Nm_{s_{k}}\!\!-\!\!Nm_{k}, 1\!\!-\!\!Nm_{k}\\
0, 1\!\!-\!\!Nm_{k}
\end{array}\right).
\end{split}
\end{equation}
\hrulefill
\end{figure*}
\begin{figure*}
\begin{equation}
\label{eq:MGF-t}
\begin{split}
\mathcal{M}_{h}(t)\!=\!\left[\prod_{k=1}^{K} \frac{L_{k}^{\alpha}}{\Gamma\left(Nm_{k}\right) \Gamma\left(Nm_{s_{k}}\right)}(\frac{\Lambda_{k}}{t})^{Nm_{k}}\right]\left(\frac{1}{2 \pi j}\right)^{K} \!\!\int_{\mathbb{C}_{1}} \!\!\int_{\mathbb{C}_{2}} \cdots\int_{\mathbb{C}_{K}}\prod_{k=1}^{K} \left\{\Gamma\left(\!-\!s_{k}\right) \Gamma\left(Nm_{s_{k}}\!+\!Nm_{k}\!+\!s_{k}\right) \Gamma\left(Nm_{k}\!+\!s_{k}\right)\right\}\\
\times\left(\frac{\Lambda_{k}}{t}\right)^{s_{k}} d s_{1} d s_{2} \ldots d s_{K}
\end{split}
\end{equation}
\hrulefill
\end{figure*}

Next, the PDF of $h$ can be obtained using the inverse Laplace transform, namely
\begin{eqnarray}
f(h)=\frac{1}{2 \pi j} \int_{\mathbb{C}} \mathcal{M}_{h}(t) e^{h t} d t.
\label{eq:inverse-Laplace}
\end{eqnarray}
\begin{figure*}
\begin{equation}
\label{eq:inverse-h-pdf}
\begin{split}
\vspace{0.1cm}f(h)&\overset{(a)}{=}\left[\prod_{k=1}^{K} \frac{L_{k}^{\alpha}\Lambda_{k}^{Nm_{k}}}{\Gamma\left(Nm_{k}\right) \Gamma\left(Nm_{s_{k}}\right)}\right]\left(\frac{1}{2 \pi j}\right)^{K} \!\int_{\mathbb{C}_{1}} \!\int_{\mathbb{C}_{2}} \!\cdots \int_{\mathbb{C}_{K}}\prod_{k=1}^{K} \left\{\Gamma\left(\!-\!s_{k}\right) \Gamma\left(Nm_{s_{k}}\!+\!Nm_{k}\!+\!s_{k}\right) \Gamma\left(Nm_{k}\!+\!s_{k}\right)\right\}\\
&\hspace{8cm}\times \Lambda_{k}^{s_{k}}\underbrace{\left(\frac{1}{2 \pi j} \int_{\mathbb{C}} t^{-\sum_{k=1}^{K}(Nm_{k}+s_{k})} e^{ht} d t\right)}_{\mathcal{I}_{1}} d s_{1} d s_{2} \ldots d s_{K}
\\
\vspace{0.1cm}&\overset{(b)}{=}\left[\prod_{k=1}^{K} \frac{L_{k}^{\alpha}\Lambda_{k}^{Nm_{k}}}{\Gamma\left(Nm_{k}\right) \Gamma\left(Nm_{s_{k}}\right)}\right]\!h^{(\sum\limits_{k=1}^{K}Nm_{k})-1}\!\left(\frac{1}{2 \pi j}\right)^{K} \!\!\int_{\mathbb{C}_{1}}\!\! \int_{\mathbb{C}_{2}}\!\!\cdots\!\!\int_{\mathbb{C}_{K}}\!\!\frac{\prod_{k=1}^{K}\left\{\Gamma\left(Nm_{k}\!+\!s_{k}\right)\Gamma\left(Nm_{s_{k}}\!+\!Nm_{k}\!+\!s_{k}\right) \right\}}{\Gamma\big(\sum_{k=1}^{K}(Nm_{k}\!+\!s_{k})\big)}\\
\vspace{0.1cm}&\hspace{10cm}\times \left\{\prod_{k=1}^{K}\Gamma\left(\!-\!s_{k}\right)\left(\Lambda_{k}h\right)^{s_{k}}\right\}d s_{1} d s_{2} \ldots d s_{K}
\\
\vspace{0.1cm}&\overset{(c)}{=}\frac{h^{(\sum\limits_{k=1}^{K}Nm_{k})-1}}{\Gamma(\sum\limits_{k=1}^{K}Nm_{k})}
\left[\prod_{k=1}^{K}(L_{k}^{\alpha}\Lambda_{k}^{Nm_{k}})\frac{\Gamma(N(m_{s_{k}}+m_{k}))}
{\Gamma(Nm_{s_{k}})}\right]\times F_{B}^{(K)}\left(N m_{1}\!+\!N m_{s_{1}}, N m_{2}\!+\!N m_{s_{2}}, \cdots, N m_{K}\!+\!N m_{s_{K}},\right.\\
\vspace{0.1cm}&\hspace{5cm}\left.N m_{1}, N m_{2},\cdots, N m_{K}; \sum\limits_{k=1}^{K}N m_{k}; -\Lambda_{1} h, -\Lambda_{2} h, \cdots, -\Lambda_{K} h\right), h \geq 0
\\
\vspace{0.2cm}&\overset{(d)}{=}\frac{h^{KNm-1}\prod_{k=1}^{K}L_{k}^{\alpha}}{B\left(KNm, KNm_{s}\right)}(\frac{\Lambda}{K})^{Nm_{k}}
\times{}_{2} F_{1}\left( KN(m+m_{s}), KNm ; KNm; -\frac{\Lambda h}{K}\right)\\
\vspace{0.2cm}&\overset{(e)}{=}\frac{\prod_{k=1}^{K}L_{k}^{\alpha}}{\Gamma(KNm)\Gamma(KNm_{s})} \dfrac{\Lambda }{K}G_{2,2}^{1,2}\left(\dfrac{\Lambda h}{K} \bigg| \begin{array}{c}
-KNm_{s},0 \\
KNm-1,0
\end{array}\right).
\end{split}
\end{equation}
\hrulefill
\end{figure*}

The derivation of $f(h)$ is given by Eq.~\eqref{eq:inverse-h-pdf}. First, substituting Eq.~\eqref{eq:MGF-t} into Eq.~\eqref{eq:inverse-Laplace}, the PDF of $h$ is rewritten at the right hand of $(a)$ in Eq.~\eqref{eq:inverse-h-pdf}. Then, assuming $y=-ht$, we solve the integral $\mathcal{I}_{1}$ with $ \frac{1}{\Gamma(\zeta)}=\frac{i}{2 \pi} \int_{C}(-t)^{-\zeta} e^{-t} d t $. $f(h)$ follows the expression at the right hand of $(b)$ in Eq.~\eqref{eq:inverse-h-pdf}, which can be represented in terms of Lauricella multivariate hypergeometric function as the expression at the right hand of $(c)$ in Eq.~\eqref{eq:inverse-h-pdf}. For the case of independent and identically distributed (i.i.d) Fisher-Snedecor RVs, the Lauricella multivariate hypergeometric function is reduced to the expression at the right hand of $(d)$ in Eq.~\eqref{eq:inverse-h-pdf}. Finally, according to Eq.~\eqref{F-function}, $f(h)$ is derived at the right hand of $(e)$ in Eq.~\eqref{eq:inverse-h-pdf}. Thus, Lemma 2 follows. $\hfill\blacksquare$

\bibliographystyle{IEEEbib}
\bibliography{References}

\end{document}